\documentclass[%
 reprint,
superscriptaddress,
 amsmath,amssymb,
 aps,
]{revtex4-2}

\usepackage{graphicx}
\usepackage{dcolumn}
\usepackage{hyperref} 
\usepackage{bm}
\usepackage{dsfont}
\usepackage{xcolor}
\usepackage{hhline} 

\newcommand{\xUZH}{Department of Physics, University of Z{\"u}rich, 8057 Z{\"u}rich, Switzerland}
\newcommand{\xcornell}{Department of Physics, Cornell University, Ithaca, NY, USA}

\begin{document}

\preprint{APS/123-QED}

\title{How quantum fluctuations freeze a classical liquid and 
then\\ melt it into a topological one}

\author{Hao Chen}
\affiliation{\xUZH}

\author{Dan Mao}
\affiliation{\xUZH}
\affiliation{\xcornell}

\author{Andrea Kouta Dagnino}
\affiliation{\xUZH}

\author{Glenn Wagner}
\affiliation{\xUZH}
\affiliation{Institute for Theoretical Physics, ETH Z{\"u}rich, 8093 Z{\"u}rich, Switzerland}

\author{Mark H. Fischer}
\affiliation{\xUZH}

\author{Juraj Hasik}
\affiliation{\xUZH}

\author{Eun-Ah Kim}
\affiliation{\xcornell}
\affiliation{Department of Physics, Ewha Womans University, Seoul, South Korea}

\author{Titus Neupert}
\affiliation{\xUZH}





\date{\today}

\begin{abstract}
Topologically ordered quantum liquids are highly sought-after quantum phases of matter, and recently, fractional Chern insulators (FCIs) joined the few experimental realizations of such phases. Here, we ask whether a gapped classical, highly degenerate liquid can be the birthplace of FCIs upon the addition of suitable quantum fluctuations. Two competing tendencies can be anticipated: (i) following the quantum order-by-disorder paradigm, quantum fluctuations could induce symmetry-breaking (charge) order, or (ii) the classical liquid builds up long-range entanglement and turns into a quantum liquid. We study spinless fermions on a honeycomb lattice subject to cluster-charging interactions and introduce quantumness through a Haldane kinetic term, featuring complex second-nearest-neighbor hopping. Based on extensive exact diagonalization calculations and high-order perturbation theory, we find that neither scenario (i) nor (ii) prevails, but (i) and (ii) manifest sequentially as the kinetic energy is increased.  We demonstrate how the gradual lifting of kinematic constraints gives rise to this sequence of phases. Our results relate to the regime of intermediate-scale interactions present in moiré systems, where band projections are not suitable to model FCIs and competing charge-ordered phases have been identified. 

\end{abstract}

\maketitle

{\it Introduction. --} The emergence of degenerate many-body ground states  in topologically ordered systems stands as one of the most striking manifestations of interactions, as nature abhors degeneracy. In the fractional quantum Hall effect at filling fraction $1/m$, the $m$-fold degenerate state is born out of a Landau level, a degenerate single-particle kinetic energy spectrum induced by the magnetic field. Weak interactions relative to the gap between the Landau levels separate the $m$ states from the extensively degenerate \textit{noninteracting} spectrum. 
In contrast, in a geometrically frustrated system, strong interactions can result in an extensively degenerate {\it many-body} spectrum in a classical limit. A kinetic energy adds quantum fluctuations to such degenerate many-body states, can lift the degeneracy through so-called ``order-by-disorder", and result in a crystalline state, such as a valence bond solid. However, quantum fluctuations can also build up long-range entanglement and result in a quantum-liquid phase with topological degeneracy, such as a spin liquid.
While the fractional Chern insulators (FCIs) that exhibit quantized Chern numbers at fractional filling of lattice systems~\cite{Tang11,Neupert2011Phys.Rev.Lett., Sun11, Regnault2011Phys.Rev.X,Wang2011Phys.Rev.Lett., Wang2012PhysRevLett.108.126805,Wu2012PhysRevB.85.075116, Lauchli2013,Liu2013PhysRevB.87.035306,
Sheng2011NatCommun,Kourtis2014Phys.Rev.Lett.,Grushin2015Phys.Rev.B, morales2024magic, Yu2024PhysRevB.109.045147,Parameswaran2012PhysRevB.85.241308,Kourtis2014Phys.Rev.Lett.a,Claassen2015PhysRevLett.114.236802,Wang2021PhysRevLett.127.246403, Varjas2022SciPostPhys.12.4.118,Abouelkomsan2023PhysRevResearch.5.L012015,Kourtis2012Phys.Rev.B,Parker2021a, Wilhelm2021Phys.Rev.Bb, Crepel2023PhysRevB.107.L201109,Lu2024PhysRevLett.133.186602} are often thought of as an extension of the fractional quantum Hall effect, the existence of a lattice necessarily opens the  door to effects of lattice geometry. 
When an FCI emerges in a lattice with three-fold rotational symmetry---the source of geometric frustration---a natural question arises: can the FCI originate from a degenerate many-body spectrum driven by this geometric frustration?

Recent experimental advances in moiré materials discovered the emergence of FCIs in small or zero external magnetic fields \cite{Zeng2023Nature, cai2023signatures,ji2024local, Xie2021Naturea,Aronson2024}, often accompanied by competing incompressible charge-ordered states with zero Chern number~\cite{Xie2021Naturea, Aronson2024}. 
Since the moiré materials have (super-) lattices with three-fold rotation symmetry, the question about the role of geometric frustration in the emergence of an FCI and its competition with charge order is more timely than ever. 

In the widely pursued weak-coupling approaches, partially filled isolated bands that resemble Landau levels go from a metal to an FCI upon adding interactions that are small compared to the band gap~\cite{Parameswaran2012PhysRevB.85.241308,Kourtis2014Phys.Rev.Lett.a,Claassen2015PhysRevLett.114.236802, Wang2021PhysRevLett.127.246403, Varjas2022SciPostPhys.12.4.118,Abouelkomsan2023PhysRevResearch.5.L012015,Kourtis2012Phys.Rev.B,Parker2021a, Wilhelm2021Phys.Rev.Bb, Crepel2023PhysRevB.107.L201109}. While this framework can also find a nesting-driven charge-density-wave state~\cite{Kourtis2012Phys.Rev.B,Parker2021a, Wilhelm2021Phys.Rev.Bb, Crepel2023PhysRevB.107.L201109}, such a charge-density-wave state is in general compressible, as only small portions of the Fermi surface acquire a gap. Most importantly, both the FCI and the charge-ordered state require fine tuning of the band structure. Alternatively, strongly interacting electrons, which partially fill a lattice, could go from a trivial compressible insulator~\cite{Sheng2011NatCommun,Kourtis2014Phys.Rev.Lett., Grushin2015Phys.Rev.B} or a classical charge-ordered state~\cite{Wang2011Phys.Rev.Lett.,Wang2012PhysRevLett.108.126805}  to an FCI upon adding complex hoppings. 
In these strong-coupling scenarios, the interaction can be of the order of the band gap and the FCI is insensitive to the details of the bandstructure. However, the insulating state in the limit of infinite interactions is either a degenerate compressible state or a classical charge-ordered state. Hence, the existing approaches from either a weak-coupling or strong-coupling angle fail to capture the competition between an incompressible \emph{quantum} charge-ordered state and an FCI. Moreover, the potential role of geometric frustration has been overlooked. 

We investigate the competition between trimer liquid (TL)~\cite{zhang2024}, generalized Wigner crystal (gWC), and FCI in a honeycomb lattice with strong cluster-charging interactions at a filling of 1/3 electrons per unit cell. The cluster-charging interaction is proposed to capture the leading-order repulsive interaction at this filling for twisted bilayer graphene~\cite{po2018origin,Kang2019PhysRevLett.122.246401, zhang2019twisted,Astrakhantsev2023}. 
These interactions introduce orbital geometric frustration, placing the system between the short-range- and the long-range-interaction limits. At filling 1/3 and infinite cluster charging, the degeneracy is extensive \cite{verberkmoes1999triangular,zhang2022fractional}, in contrast to the Coulomb interaction, where a gWC is the ground state. However, despite the extensive degeneracy, the state is incompressible, contrary to the nearest-neighbor-repulsion case \cite{Kourtis2014Phys.Rev.Lett., Grushin2015Phys.Rev.B}. 

The Haldane hopping terms introduce quantum fluctuations with two possible competing tendencies. On the one hand, quantum fluctuations can generate ordering, such as in the valence bond solids, via quantum order-by-disorder in dimer models on bipartite lattices~\cite{RK,moessner2001short}. On the other hand, the Haldane hopping terms imprint the band topology of Chern bands, which may favor an FCI. Using exact diagonalization (ED) and high-order perturbation theory in the strong-coupling limit, we show that quantum fluctuations play a dual role: 1. For small hopping, quantum fluctuations lift the degeneracy of the cluster-charging term and favor a gWC via fourth-order virtual hopping processes. 2. An FCI phase is realized in the intermediate hopping regime, where the bandwidth of the Chern band and the band gap are comparable to the interaction. 

\begin{figure}[tb]
    \centering
    \includegraphics[width=\columnwidth]{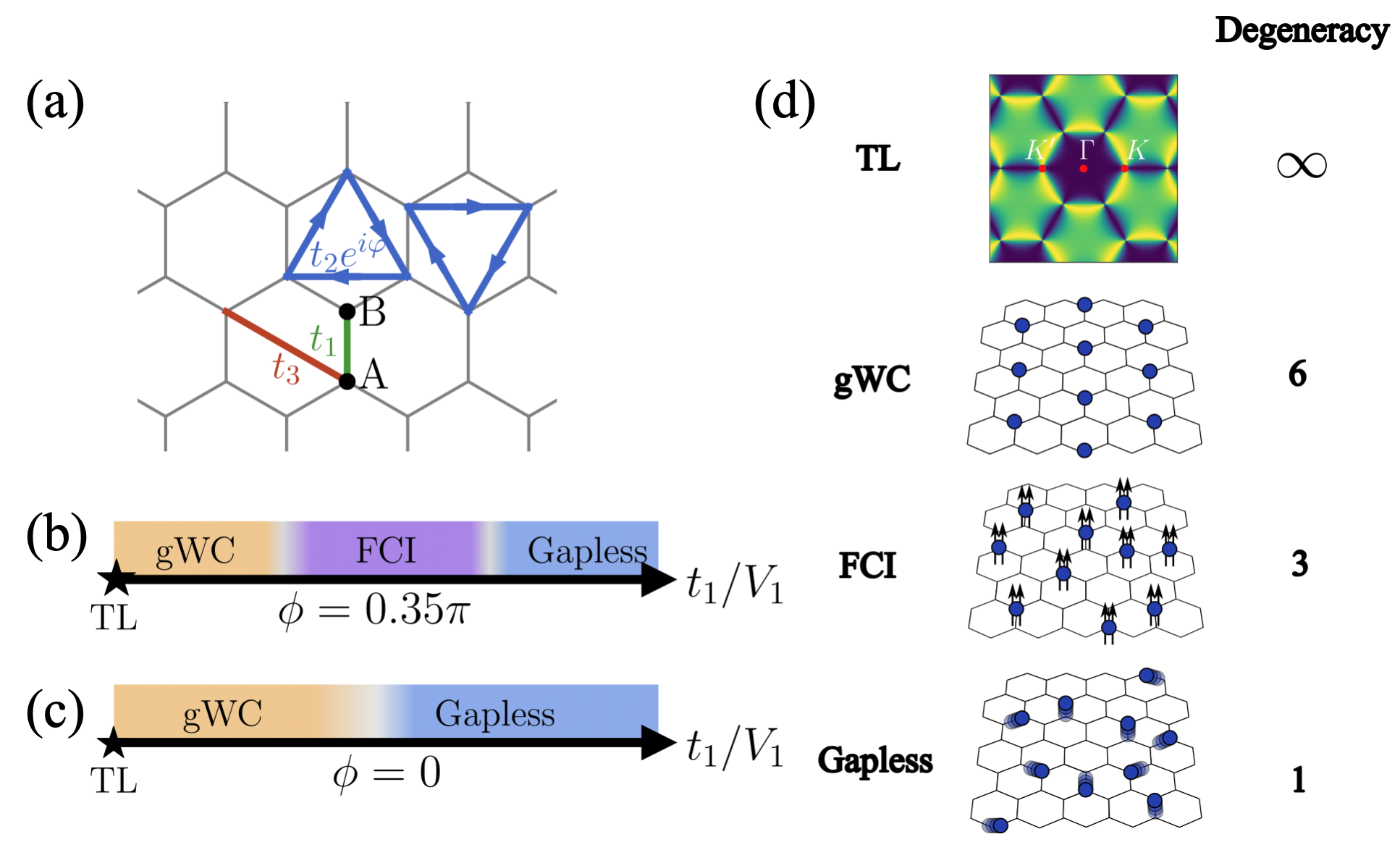}
    \caption{(a) Haldane model on a honeycomb lattice. Possible hopping terms are depicted in different colors, with the arrows for 2NN hopping indicating the phase convention.
    Qualitative phase diagrams of the model as a function of $t_1/V_1$ for $\phi = 0.35\pi$ (b) and $\phi=0$ (c), respectively. TL stands for the trimer liquid phase
    in the infinite-interaction limit $t_1/V_1=0$. gWC denotes generalized Wigner crystal and FCI denotes fractional Chern insulator. (d) Characteristics/cartoons and degeneracies of the phases in (b) and (c). TL features two two-fold pinch points at $\mathbf{K}$ and $\mathbf{K^\prime}$ of the static structure factor in momentum space, reflecting the constraint imposed by the cluster-charging interaction.
    }
    \label{fig:model_schematic}
\end{figure}

{\it Model. --}
We study the cluster-charging Hamiltonian with the Haldane-model hopping for spinless fermions,
\begin{equation}\label{eq:full_Hamiltonian}
    \hat{H} = -\sum_{i,j} t_{ij} \hat{c}_i^\dagger \hat{c}_j
        + V_1 \sum_{\langle ij \rangle_1} \hat{n}_i \hat{n}_j
        + V_2 \sum_{\langle ij \rangle_2} \hat{n}_i \hat{n}_j
        + V_3 \sum_{\langle ij \rangle_3} \hat{n}_i \hat{n}_j,
\end{equation}
where $\langle ij \rangle_1, \langle ij \rangle_2$, and $\langle ij \rangle_3$ denote nearest neighbors (NN), second nearest neighbors (2NN), and third nearestneighbors (3NN), respectively. The cluster-charging interaction is realized when $V_1 = 2V_2 =2V_3$. Without the hopping terms, the ground state of the Hamiltonian is the so-called trimer-liquid phase of Ref.~\onlinecite{zhang2024}, with extensive degeneracy. Fermions can hop between NN, 2NN, and 3NN with hopping amplitudes $t_{\langle ij \rangle_1}=t_1, t_{\langle ij \rangle_2}=t_2 e^{\pm i\phi},$ and $t_{\langle ij \rangle_3} = t_3$, respectively. The sign convention of the acquired phase $\pm \phi$ is illustrated in Fig.~\ref{fig:model_schematic}(a). 
We perform an ED study on the  model given in Eq.~\eqref{eq:full_Hamiltonian}, with the following choice for the hopping parameters: $t_2/t_1 = 0.7, t_3/t_1 = -0.9$, and $\phi=0.35\pi$ or $\phi=0$ as specified below. Under this choice, the non-interacting part of the Hamiltonian features two Chern bands with opposite (single-particle) Chern numbers $C=\pm 1$. 

{\it Exact Diagonalization. --} In ED, we consider finite clusters consisting of $N_1\times N_2$ unit cells (with $N_{\text{s}} = 2N_1 N_2$ the total number of sites) under periodic boundary conditions and exploit tilted geometries \cite{Lauchli2013,Repellin2014,si} to achieve near-unity aspect ratios.
The cluster geometries used in this work are illustrated in Fig.~\ref{fig:all_sizes}, with the corresponding parameters provided in Appendix~\ref{app:Tilted_BC}.
We focus on the filling fraction $\nu = N_{e}/(N_1N_2) =1/3$.

Figures~\ref{fig:model_schematic}(b) and (c) present the qualitative phase diagrams with $\phi=0.35\pi$ and $\phi=0$, respectively. The horizontal axis represents $t_1/V_1$, which is gradually increased, along with the corresponding adjustments to $t_2$ and $t_3$. 
In both cases, the model forms a critical trimer liquid at $t_1/V_1=0$ \cite{verberkmoes1999triangular,Mao2023,zhang2024}.
For small $t_1/V_1$, a gWC phase with a $\sqrt{3} \times \sqrt{3}$ charge order emerges, featuring six degenerate ground states. The number six originates from the two-fold degeneracy from the sublattice symmetry times three-fold degeneracy from tripling of the unit cell.
In the opposite limit, $t_1/V_1\rightarrow \infty$, a gapless Fermi liquid phase forms. Unlike the trivial hopping phase $\phi=0$, an FCI with a three-fold degeneracy (on a torus) emerges for $\phi=0.35\pi$.

\begin{figure}[t]
    \centering
    \includegraphics[width=\columnwidth]{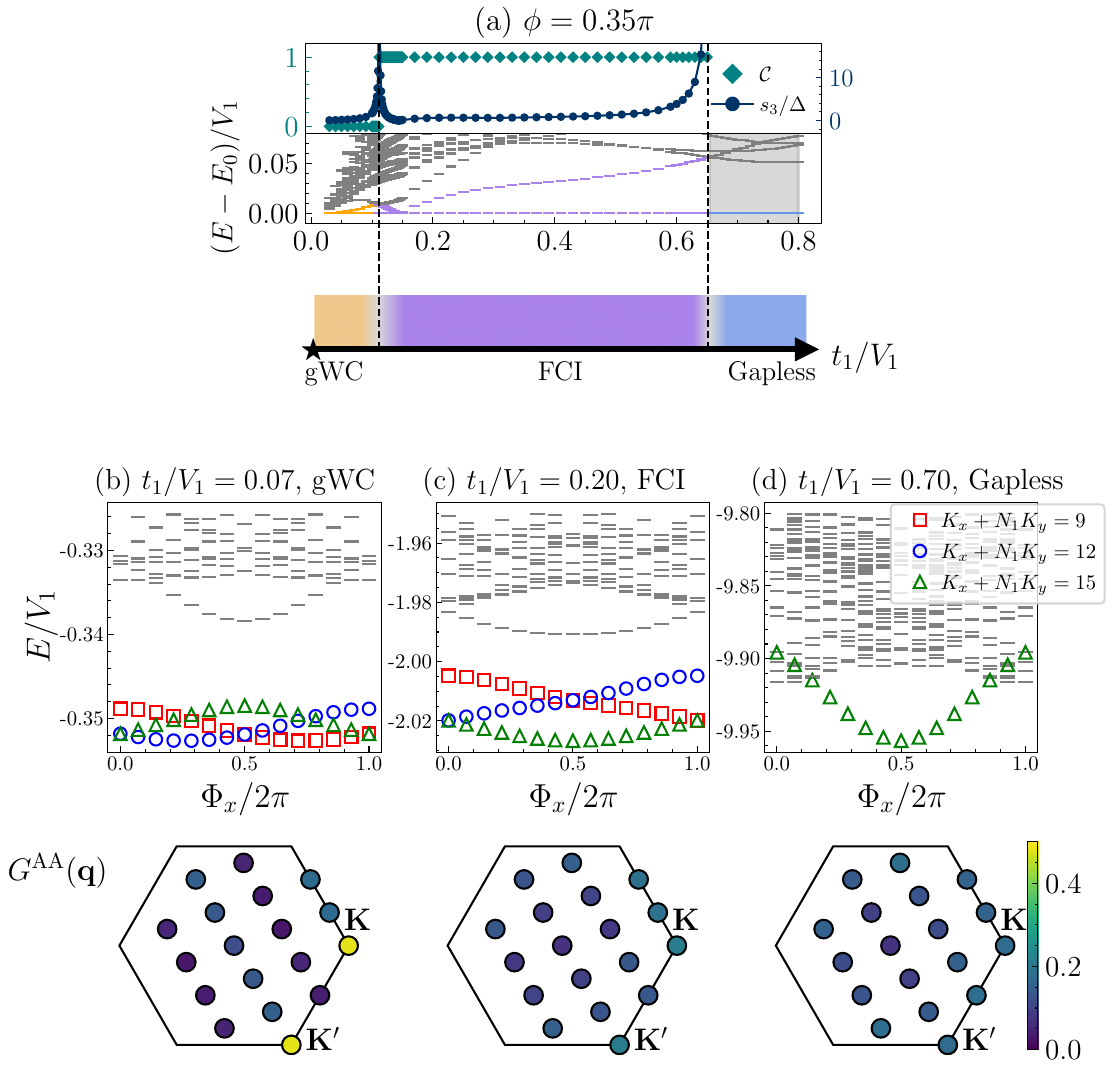}
    \caption{Characterizations of the phases of Eq.~\eqref{eq:full_Hamiltonian} on the $9\times 2$ cluster.
    (a) Evolution of the low-lying spectrum as $t_1/V_1$ increases. In the upper plot we show the many-body Chern number $\mathcal{C}$ and the spacing/gap ratio $s_3/\Delta$ for the ground states of the gapped phases.
    The spectral flows and ground state structure factors $G^{\text{AA}}(\mathbf{q})$ of the three phases are shown in panels (b), (c) and (d). The structure factor is calculated for the ground state with momentum number $K_x + N_1K_y=15$.
    In the gWC and FCI phases, the ground states occupy the same momentum sectors, marked in non-grey colors. Both phases exhibit similar spectral flow patterns: the three ground states flow into each other upon flux insertion, maintaining a persistent gap from the excited states.  This spectral flow pattern is to be compared with the level crossings shown in the gapless phase (panel (c)).
    The structure factor of the gWC phase is peaked at $\mathbf{K}$ and $\mathbf{K}^\prime$, indicating a translational-symmetry-breaking charge order with a tripled unit cell.}
    \label{fig:spec_flow&corr}
\end{figure}

Figure~\ref{fig:spec_flow&corr} presents numerical evidence for the three phases at $\phi=0.35\pi$, using the $9\times2$ cluster as an example. The evolution of the low-lying spectrum when varying $t_1/V_1$ is shown in Fig.~\ref{fig:spec_flow&corr}(a). We found two gapped phases separated by a gap-closing point. Both of them exhibit a three-fold degeneracy on this cluster. 
The mismatch between the observed three-fold degeneracy in the small $t_1/V_1$ regime on the finite cluster and the six-fold degeneracy of the $\sqrt{3}\times\sqrt{3}$ phase is due to finite size effects: the three-fold degeneracy observed on the finite cluster comes from the tripling of unit cells, but the finite-size effects lift the two-fold degeneracy between two sublattices.
The gap closing can be seen from the divergence of spacing/gap ratio $s_3/\Delta$, where $s_3$ is the maximum spacing between adjacent levels in the three-fold degenerate ground states and $\Delta=E_4-E_3$ is the difference between the fourth-lowest energy $E_4$ and 
the third-lowest energy $E_3$. Further increasing $t_1/V_1$ drives the system into a gapless state (shaded area).
In determining whether the ground state is potentially gapless, we introduce twisted boundary conditions along the two periodic vectors $(\mathbf{T}_1, \mathbf{T}_2)$ of the cluster. Specifically, we enforce particles to acquire an additional phase $\Phi_x (\Phi_y)$ when translated by $\mathbf{T}_1 (\mathbf{T_2}$).
We consider a ground state as gapped only if the many-body gap persists for all possible flux configurations $(\Phi_x, \Phi_y) \in [0, 2\pi)^{\times 2}$. Note that Fig.~\ref{fig:spec_flow&corr}(a) only shows the spectrum evolution at a specific flux configuration $(\Phi_x, \Phi_y) = (\pi, 0)$, for which the gap-closing point between the two gapped phases can be directly observed. The seemingly non-zero gap in the gapless region is due to the flux-configuration choice. 
In Fig.~\ref{fig:spec_flow&corr}(b-d) we present the spectral flow upon varying $\Phi_x$ for the three phases. The level crossings in Fig.~\ref{fig:spec_flow&corr} (d) clearly demonstrates the gapless nature in the large-$t_1/V_1$ regime, while a persistent gap is found for the two gapped phases.
However, the spectral flows provide limited insight on distinguishing the two gapped phases in our case, as the ground states of both phases lie in the same set of momentum sectors and exhibit similar spectral flows for all clusters studied in this work, as illustrated in Fig.~\ref{fig:spec_flow&corr}(b) and (c).

To further characterize the two gapped phases, we measure the ground state structure factor $G_i^{\text{AA}}(\mathbf{q})$ between the A-A sublattice sites, defined as
\begin{equation}
\begin{aligned}
    G_i^{\text{AA}}(\boldsymbol{q}) &= 
    \frac{1}{N_1N_2}\left(\langle g_i| \hat{n}^{\text{A}}_{\mathbf{q}} \hat{n}^{\text{A}}_{\mathbf{-q}}
    |g_i\rangle - \langle g_i | \hat{n}^{\text{A}}_{\mathbf{q}=0} | g_i \rangle^2 \delta_{\mathbf{q}, 0}\right),\\
\end{aligned}
\end{equation}
where $|g_i\rangle$ is one of the ground states and 
$\hat{n}^{\text{A}}_{\mathbf{q}} = \sum_{\mathbf{r}_j} \hat{n}^{\text{A}}_{\mathbf{r}_j} e^{i\mathbf{q}\cdot \mathbf{r}_j}$.
The structure factors $G^{\text{AA}}(\mathbf{q})$ in Fig.~\ref{fig:spec_flow&corr}(b) and (c) reveal distinct characteristics:
the small-$t_1/V_1$ phase displays a charge order with ordering momenta $\mathbf{K}$ and $\mathbf{K}^\prime$, 
whereas the second phase exhibits a roughly uniform $G^{\text{AA}}(\mathbf{q})$, indicating no translational symmetry breaking.
Furthermore, we compute the many-body Chern number $\mathcal{C}$ for the three-fold degenerate ground states,
defined as a topological invariant obtained through integration over the twisted boundary conditions, as detailed in Appendix~\ref{app:chern}).
The Hall conductivity $\sigma_{xy}$ is related to the many-body Chern number by $\sigma_{xy} = \frac{1}{m} \frac{\mathcal{C} e^2}{h}$, where $m=3$ is the degeneracy of the ground states. 
The Chern-number calculations demonstrate that the small-$t_1/V_1$ phase is a symmetry-breaking phase with $\mathcal{C}=0$, whereas
the second phase is identified as an FCI state with $\mathcal{C}=1$ (see Appendix~\ref{app:chern}), as shown in the upper plot of Fig.~\ref{fig:spec_flow&corr}(a). 


In Fig.~\ref{fig:all_sizes}, we plot the many-body gap $\Delta$ as a function of $t_1/V_1$ for all clusters studied in ED. 
Here, we use the notation $(N_1, N_2, \text{Deg})$ to distinguish tilted clusters with the same $(N_1, N_2)$, where Deg refers to the classical ground state degeneracy at $t_1/V_1 = 0$. For the $(9,2,48)$ cluster, $\Delta$ is determined by the minimum gap among all possible flux configurations $(\Phi_x, \Phi_y)$. For larger clusters, only $\Delta$ at the fixed $(\Phi_x, \Phi_y) = (0,0)$ is plotted, which is sufficient to indicate the gap closing at the phase transition. We expect the variation of $\Delta$ with $(\Phi_x, \Phi_y)$ to be suppressed for larger system sizes,
as the flux-induced vector potential $A_i \sim \Phi_{i}/L_{i} \to 0$ vanishes in the thermodynamic limit.
Among the five clusters, three (colored blue) exhibit the same phase structure as previously observed: a transition from a gWC phase to an FCI phase and finally to a gapless phase, as confirmed by Chern-number calculations. However, the other two clusters (colored red) show a gapless spectrum in the small-$t_1/V_1$ regime, despite the geometries being compatible with the tripled unit cell structure expected for the gWC phase. 
Irrespectively, as $t_1$ increases, all clusters invariably enter into an FCI phase, which suggests that the FCI phase is robust and likely to persist in the thermodynamic limit.
Motivated by the inconsistency in the small-$t_1/V_1$ regime, in the following section we develop a perturbative analysis to address finite-size effects.

\begin{figure}[t]
    \centering
    \includegraphics[width=\columnwidth]{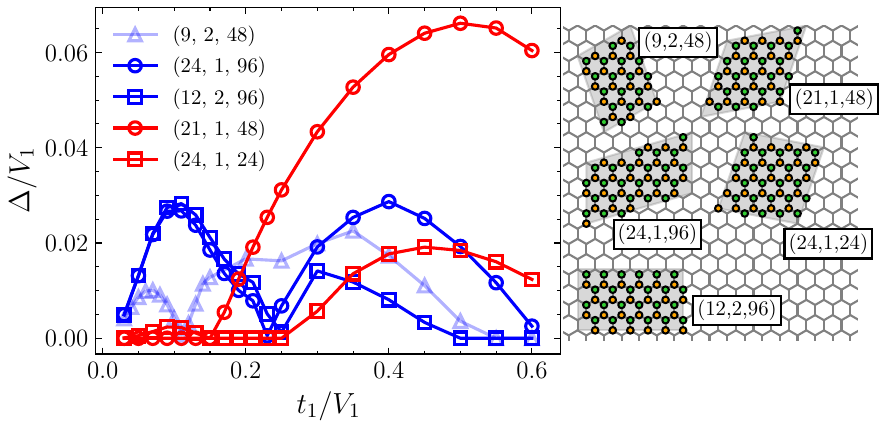}
    \caption{Many-body gap $\Delta=E_4 - E_3$ as a function of $t_1/V_1$ for finite clusters. 
    Clusters are labeled by $(N_1, N_2, \text{Deg})$ with Deg the classical ground state degeneracy at $t_1/V_1 = 0$. The geometries of the clusters are shown in the right panel.}
    \label{fig:all_sizes}
\end{figure}

{\it Perturbative analysis. --} To better understand the charge ordering in the limit of $t_1/V_1\to0$, we perform a perturbative analysis. We write the Hamiltonian of Eq.~\eqref{eq:full_Hamiltonian} as $\hat{H} = \hat{H}_0 + \hat{H}_1$ with $\hat{H}_0$ the cluster-charging interaction and  $\hat{H}_1 = -\sum_{i, j}t_{ij} \hat{c}_i^\dagger \hat{c}_j$ as the perturbation. The eigenstates of $H_0$ form degenerate flat bands with energies $E_{n}^{(0)} = nV_1/2$, and we use $\hat{P}$ to denote the projector onto the ground state manifold of $\hat{H}_0$.
By applying a Schrieffer-Wolff (SW) transformation \cite{si, Bravyi2011, Slagle2017}, we obtain the effective Hamiltonian defined on the ground state manifold of $\hat{H}_0$:
\begin{equation}
    \hat{H}_{\text{eff}} = \sum_{n=1}^{\infty} \hat{P} (\hat{H}_1 \hat{D})^{n-1} \hat{H}_1 \hat{P} + \cdots,
\end{equation}
where $\hat{D} = (\mathds{1}-\hat{P})/(\hat{H}_0 - E_0^{(0)})$, which projects the intermediate states onto the high-energy space of $\hat{H}_0$, and $\cdots$ represents the ``disconnected" terms such as $\hat{P} \hat{H}_1 \hat{D} \hat{H}_1 \hat{P} \hat{H}_1 \hat{D}^2 \hat{H}_1 \hat{P}$.
When evaluated in the occupation number basis $|n_1, \cdots, n_{N_s} \rangle$, the effective Hamiltonian can be interpreted as a set of virtual hopping processes, which connect one classical ground state to another.  
Importantly, under periodic boundary conditions, there exist virtual processes in which the hopping electrons wrap around the torus. 

On finite clusters, the order of these virtual hopping processes is proportional to the linear system size $L$, and they vanish as $(t_1/V_1)^{\alpha L}$ with $\alpha \sim O(1)$ when $L\to \infty$. However, on small clusters they can dominate over the true leading-order processes that survive in the thermodynamic limit, explaining the inconsistency  in the small-$t_1/V_1$ regime.
To avoid the dynamics dominated by such spurious terms, we directly calculate the effective Hamiltonian in the thermodynamic limit. For the off-diagonal part of $\hat{H}_{\text{eff}}$, it has been shown that the lowest-order nontrivial process is at the eighth order, giving rise to the lemniscate operator \cite{Mao2023}. The diagonal part of $\hat{H}_{\text{eff}}$, in contrast, can have lower-order contributions, which have the form
\begin{equation}
    \hat{H}_{\text{eff}}^{\text{diag}} = \sum_{ij} V_{ij} \hat{n}_i \hat{n}_j + \sum_{ijk} V_{ijk} \hat{n}_i \hat{n}_j \hat{n}_k + \cdots.
\end{equation}

\begin{figure}[t]
    \centering
    \includegraphics[width=0.8\columnwidth]{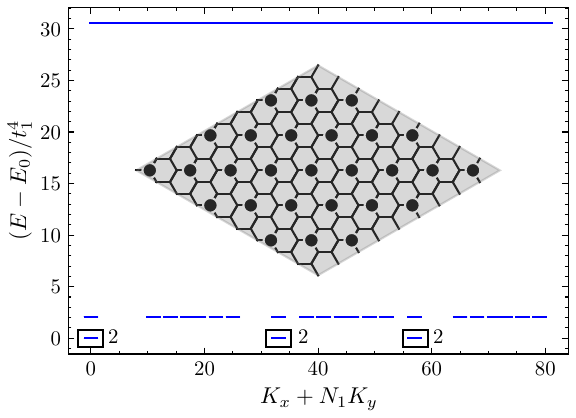}
    \caption{Low-lying spectrum of the effective Hamiltonian $\hat{H}_{\text{eff}}^{(4)}$ up to fourth order on a $9\times9$ cluster. Within each box the energy level is doubly degenerate. There are in total six degenerate ground states. The inset illustrates the fermion number distribution for one of the ground states, where the fermions are colored in black.}.
    \label{fig:eff_spec_9x9}
\end{figure}
Up to third order, all ground-state configurations are renormalized by the same energy, while non-trivial splittings among the ground states appear for fourth-order diagonal terms. As listed in the Appendix~\ref{app:perturbation}, the fourth-order diagonal terms only involve a (two-body) density-density interaction, ranging from 4NN to 13NN. Therefore, in the limit of $t_1/V_1 \ll 1$, the off-diagonal lemniscate operator is sub-leading compared to the diagonal terms.  
With $\hat{H}_{\text{eff}}$ obtained in the thermodynamic limit, we then calculate its energy spectrum on finite clusters much larger than those accessible with full ED. Since the spurious processes responsible for finite-size effects have been excluded in the calculation of $\hat{H}_{\text{eff}}$, the resulting ground states are expected to correspond to those in the thermodynamic limit.
Figure~\ref{fig:eff_spec_9x9} shows the low-lying spectrum of $\hat{H}_{\text{eff}}^{(4)} = \sum_{i, j} V_{ij} \hat{n}_i \hat{n}_j$ on a $9\times 9$ cluster, where a six-fold degeneracy is found. 
Figure~\ref{fig:eff_spec_9x9} also illustrates the particle distribution in one of the six ground states, which demonstrates that the ground state has a $\sqrt{3}\times \sqrt{3}$ structure. 
Those configurations maximize the number of 5NN interactions per fermion. However, the competition between the 4NN, 5NN and 6NN interactions can give rise to a stripe order when the phase $\phi$ of the NNN hopping  is tuned (see Appendix~\ref{app:perturbation}).

{\it FCI from constrained Hilbert space. --}
The FCI phase of the Hamiltonian given in Eq.~\eqref{eq:full_Hamiltonian} is in a distinct regime from the FCI observed in partially filled Chern band with negligible band mixing \cite{Parameswaran2012PhysRevB.85.241308,Claassen2015PhysRevLett.114.236802, Wang2021PhysRevLett.127.246403, Varjas2022SciPostPhys.12.4.118,Abouelkomsan2023PhysRevResearch.5.L012015,Parker2021a, Wilhelm2021Phys.Rev.Bb, Crepel2023PhysRevB.107.L201109}. In the latter scenarios, the single-particle band gap is considered to be much larger than the interaction strength, allowing the physics to be fully captured by the lower Chern band, which plays the role of the lowest Landau level in the fractional quantum Hall effect. In our case, the system enters the FCI phase when the band gap of the single-particle Hamiltonian $\Delta_{\text{single}} \approx 0.6 t_1$ is comparable to the interaction strength,
making the lower band projection picture unjustified. 
Additionally, the band width of the lower Chern band is larger than the band gap $\delta \approx 0.86 t_1 > \Delta_{\text{single}}$, deviating from the ideal flat-band limit.
FCIs in comparable regimes were previously studied in Refs.~\cite{Kourtis2014Phys.Rev.Lett.a,Kourtis2014Phys.Rev.Lett.}.
To better understand how the FCI arises in this intermediate regime, we adopt a complementary approach that starts from the strong coupling limit of the cluster-charging Hamiltonian $\hat{H}_0$ and study the quantum fluctuations induced by the hopping terms. 
As the hopping amplitudes increase, more low-lying eigenstates of $\hat{H}_0$ are activated and contribute to the ground states. In this picture, the hierarchy of interaction levels provides a way to truncate the Hilbert space, with the hopping amplitudes determining how many energy levels are included in the description. Motivated by this observation, we propose the following model. First, we introduce an energy cutoff $E_{\text{cutoff}}$ and truncate the Hilbert space by keeping states with energy $E \leq E_{\text{cutoff}}$ in the classical cluster-charging model. In the truncated Hilbert space, we then discard the interactions and study the hopping Hamiltonian on this restricted Hilbert space. The final Hamiltonian is written as
\begin{equation}
    \hat{H}_{\text{c}}(E_{\text{cutoff}}) = \hat{P}(E_{\text{cutoff}}) \bigg[-\sum_{i, j} t_{ij} \hat{c}_i^\dagger \hat{c}_j\bigg] \hat{P}(E_{\text{cutoff}}),
\end{equation}
where $\hat{P}(E_{\text{cutoff}})$ denotes the projector onto the space with $E \leq E_{\text{cutoff}}$ in the classical model.

\begin{figure}[t]
    \centering
    \includegraphics[width=\columnwidth]{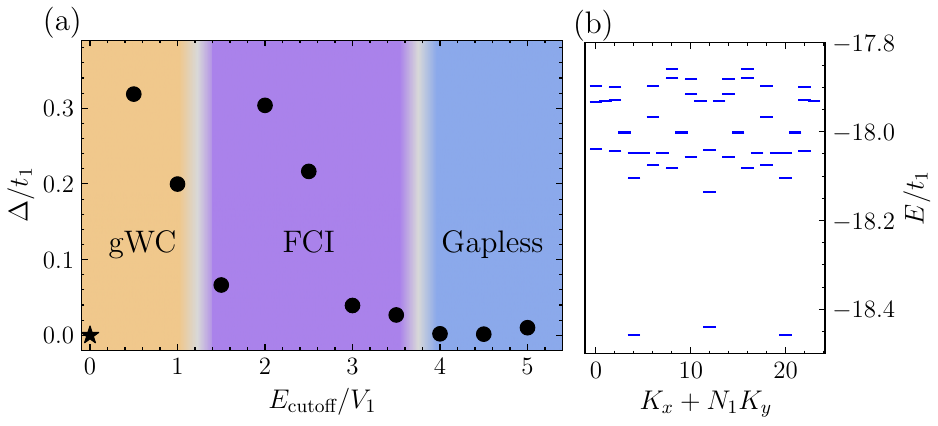}
    \caption{(a) Phase diagram of $\hat{H}_{\text{c}}$ on the $(N_1, N_2, \text{deg}) = (24, 1, 96)$ cluster, with black data points representing the gap. 
    Results are shown for $E_{\text{cutoff}} \leq 5 V_1$, but note that there are states with cluster-charging energies above $5V_1$.
    (b) The low-lying spectrum of the FCI phase at $E_{\text{cutoff}}/V_1 = 2$.}
    \label{fig:damp0_24_1}
\end{figure}

Using ED, we find that the constrained model $\hat{H}_c$ reproduces the same phase diagram as the full Hamiltonian in Eq.~\eqref{eq:full_Hamiltonian} on the finite clusters analyzed in the previous sections (see Appendix~\ref{app:constrained_model}), with $E_{\text{cutoff}}$ replacing $t_1$. As an example, we plot the gap $\Delta=E_4 - E_3$ of $\hat{H}_{\text{c}}$ as a function of $E_{\text{cutoff}}$ on the $(N_1, N_2, \text{deg}) = (24, 1, 96)$ cluster in Fig.~\ref{fig:damp0_24_1}(a). Figure~\ref{fig:damp0_24_1}(b) shows the energy spectrum of the FCI phase at $E_{\text{cutoff}} = 2V_1$, where the three-fold-degenerate ground-state space carries a non-zero Chern number $\mathcal{C}=1$, obtained from numerical calculations.
The correspondence between $\hat{H}_c$ and $\hat{H}$ indicates that the FCI phase emerges from the first several levels of the classical model. In this picture, the interaction provides a natural way to energetically truncate the Hilbert space, with the states within the sub-Hilbert space satisfying the constraint that the number of frustrated plaquettes is bounded by some fixed number. The single-particle kinetic terms introduce coupling between the states within the subspace, resulting in a topologically ordered FCI state. It contrasts with the Landau level picture, where the sub-Hilbert space is formed by the lower Chern band, and the interaction projected to the lower band leads to an FCI state.

{\it Conclusion. --} 
In summary, we studied the effect of quantum fluctuations in the geometrically frustrated setting of a strongly-interacting cluster-charging model at fractional filling of the honeycomb lattice on the torus. We found the time-reversal-breaking hopping in the form of the Haldane model to gradually reduce the exponential degeneracy of the novel TL phase first into an incompressible gWC phase with six-fold degeneracy, and then to an FCI with three-fold degeneracy, and eventually into a gapless metallic state upon increasing the strength of quantum fluctuations. 
This model provides an intriguing example of competing gWC and FCI ordering, both driven by quantum fluctuations. Notably, both phases are reached without fine-tuning of the band structure, as the interaction strength is on par with the band gap. The richness of the phase diagram we obtained showcases how the frustrated intermediate range interaction of cluster charging and the resulting gapless incompressible TL phase serve as a fertile birth place for correlated states of matter. Notably, the FCI phase we found is an example of the ``Quantum Charge Liquid" \cite{musser2024fractionalizationalternatechargeordering}, as ansalogous to the quantum spin liquid.
Our results invite the questions such as whether the FCI can be reached directly from the TL through spontaneous time-reversal-symmetry breaking as in a chiral spin liquid state, and what the nature of the transition between gWC and FCI is. Additionally, a natural question to ask is whether our model can provide insights into the competition among different phases observed in recent experiments on moir\'e systems. In moir\'e materials such as twisted MoTe$_2$, the tunable twist angle and displacement field change the kinetic energy in a complicated way. However, the mechanisms for stabilizing the gWC and the FCI seem generic and mostly dependent on the overall band width versus interaction strength. In particular, the exchange terms that favor the gWC would also be present for a generic screened Coulomb interaction, a topic that we leave for future studies.


{\it Acknowlegments. --} G.W.~is supported by the Swiss National Science Foundation (SNSF) via Ambizione grant number PZ00P2-216183. This project was supported by the Swiss National Science Foundation through a Consolidator Grant (iTQC, TMCG-2-213805). D.M. and E.-A.K. were partially supported by the Gordon and Betty Moore Foundation’s EPiQS Initiative, Grant GBMF10436 to  E.-A.K. E.-A.K. acknowledges support by the NSF through the grant OAC-2118310.

\bibliography{ref}

\begin{thebibliography}{46}%
\makeatletter
\providecommand \@ifxundefined [1]{%
 \@ifx{#1\undefined}
}%
\providecommand \@ifnum [1]{%
 \ifnum #1\expandafter \@firstoftwo
 \else \expandafter \@secondoftwo
 \fi
}%
\providecommand \@ifx [1]{%
 \ifx #1\expandafter \@firstoftwo
 \else \expandafter \@secondoftwo
 \fi
}%
\providecommand \natexlab [1]{#1}%
\providecommand \enquote  [1]{``#1''}%
\providecommand \bibnamefont  [1]{#1}%
\providecommand \bibfnamefont [1]{#1}%
\providecommand \citenamefont [1]{#1}%
\providecommand \href@noop [0]{\@secondoftwo}%
\providecommand \href [0]{\begingroup \@sanitize@url \@href}%
\providecommand \@href[1]{\@@startlink{#1}\@@href}%
\providecommand \@@href[1]{\endgroup#1\@@endlink}%
\providecommand \@sanitize@url [0]{\catcode `\\12\catcode `\$12\catcode
  `\&12\catcode `\#12\catcode `\^12\catcode `\_12\catcode `\%12\relax}%
\providecommand \@@startlink[1]{}%
\providecommand \@@endlink[0]{}%
\providecommand \url  [0]{\begingroup\@sanitize@url \@url }%
\providecommand \@url [1]{\endgroup\@href {#1}{\urlprefix }}%
\providecommand \urlprefix  [0]{URL }%
\providecommand \Eprint [0]{\href }%
\providecommand \doibase [0]{https://doi.org/}%
\providecommand \selectlanguage [0]{\@gobble}%
\providecommand \bibinfo  [0]{\@secondoftwo}%
\providecommand \bibfield  [0]{\@secondoftwo}%
\providecommand \translation [1]{[#1]}%
\providecommand \BibitemOpen [0]{}%
\providecommand \bibitemStop [0]{}%
\providecommand \bibitemNoStop [0]{.\EOS\space}%
\providecommand \EOS [0]{\spacefactor3000\relax}%
\providecommand \BibitemShut  [1]{\csname bibitem#1\endcsname}%
\let\auto@bib@innerbib\@empty
\bibitem [{\citenamefont {Tang}\ \emph {et~al.}(2011)\citenamefont {Tang},
  \citenamefont {Mei},\ and\ \citenamefont {Wen}}]{Tang11}%
  \BibitemOpen
  \bibfield  {author} {\bibinfo {author} {\bibfnamefont {E.}~\bibnamefont
  {Tang}}, \bibinfo {author} {\bibfnamefont {J.-W.}\ \bibnamefont {Mei}},\ and\
  \bibinfo {author} {\bibfnamefont {X.-G.}\ \bibnamefont {Wen}},\ }\bibfield
  {title} {\bibinfo {title} {High-temperature fractional quantum hall states},\
  }\href {https://doi.org/10.1103/PhysRevLett.106.236802} {\bibfield  {journal}
  {\bibinfo  {journal} {Phys. Rev. Lett.}\ }\textbf {\bibinfo {volume} {106}},\
  \bibinfo {pages} {236802} (\bibinfo {year} {2011})}\BibitemShut {NoStop}%
\bibitem [{\citenamefont {Neupert}\ \emph {et~al.}(2011)\citenamefont
  {Neupert}, \citenamefont {Santos}, \citenamefont {Chamon},\ and\
  \citenamefont {Mudry}}]{Neupert2011Phys.Rev.Lett.}%
  \BibitemOpen
  \bibfield  {author} {\bibinfo {author} {\bibfnamefont {T.}~\bibnamefont
  {Neupert}}, \bibinfo {author} {\bibfnamefont {L.}~\bibnamefont {Santos}},
  \bibinfo {author} {\bibfnamefont {C.}~\bibnamefont {Chamon}},\ and\ \bibinfo
  {author} {\bibfnamefont {C.}~\bibnamefont {Mudry}},\ }\bibfield  {title}
  {\bibinfo {title} {Fractional {{Quantum Hall States}} at {{Zero Magnetic
  Field}}},\ }\href {https://doi.org/10.1103/PhysRevLett.106.236804} {\bibfield
   {journal} {\bibinfo  {journal} {Physical Review Letters}\ }\textbf {\bibinfo
  {volume} {106}},\ \bibinfo {pages} {236804} (\bibinfo {year}
  {2011})}\BibitemShut {NoStop}%
\bibitem [{\citenamefont {Sun}\ \emph {et~al.}(2011)\citenamefont {Sun},
  \citenamefont {Gu}, \citenamefont {Katsura},\ and\ \citenamefont
  {Das~Sarma}}]{Sun11}%
  \BibitemOpen
  \bibfield  {author} {\bibinfo {author} {\bibfnamefont {K.}~\bibnamefont
  {Sun}}, \bibinfo {author} {\bibfnamefont {Z.}~\bibnamefont {Gu}}, \bibinfo
  {author} {\bibfnamefont {H.}~\bibnamefont {Katsura}},\ and\ \bibinfo {author}
  {\bibfnamefont {S.}~\bibnamefont {Das~Sarma}},\ }\bibfield  {title} {\bibinfo
  {title} {Nearly flatbands with nontrivial topology},\ }\href
  {https://doi.org/10.1103/PhysRevLett.106.236803} {\bibfield  {journal}
  {\bibinfo  {journal} {Phys. Rev. Lett.}\ }\textbf {\bibinfo {volume} {106}},\
  \bibinfo {pages} {236803} (\bibinfo {year} {2011})}\BibitemShut {NoStop}%
\bibitem [{\citenamefont {Regnault}\ and\ \citenamefont
  {Bernevig}(2011)}]{Regnault2011Phys.Rev.X}%
  \BibitemOpen
  \bibfield  {author} {\bibinfo {author} {\bibfnamefont {N.}~\bibnamefont
  {Regnault}}\ and\ \bibinfo {author} {\bibfnamefont {B.~A.}\ \bibnamefont
  {Bernevig}},\ }\bibfield  {title} {\bibinfo {title} {Fractional {{Chern
  Insulator}}},\ }\href {https://doi.org/10.1103/PhysRevX.1.021014} {\bibfield
  {journal} {\bibinfo  {journal} {Physical Review X}\ }\textbf {\bibinfo
  {volume} {1}},\ \bibinfo {pages} {021014} (\bibinfo {year}
  {2011})}\BibitemShut {NoStop}%
\bibitem [{\citenamefont {Wang}\ \emph {et~al.}(2011)\citenamefont {Wang},
  \citenamefont {Gu}, \citenamefont {Gong},\ and\ \citenamefont
  {Sheng}}]{Wang2011Phys.Rev.Lett.}%
  \BibitemOpen
  \bibfield  {author} {\bibinfo {author} {\bibfnamefont {Y.-F.}\ \bibnamefont
  {Wang}}, \bibinfo {author} {\bibfnamefont {Z.-C.}\ \bibnamefont {Gu}},
  \bibinfo {author} {\bibfnamefont {C.-D.}\ \bibnamefont {Gong}},\ and\
  \bibinfo {author} {\bibfnamefont {D.~N.}\ \bibnamefont {Sheng}},\ }\bibfield
  {title} {\bibinfo {title} {Fractional {{Quantum Hall Effect}} of {{Hard-Core
  Bosons}} in {{Topological Flat Bands}}},\ }\href
  {https://doi.org/10.1103/PhysRevLett.107.146803} {\bibfield  {journal}
  {\bibinfo  {journal} {Physical Review Letters}\ }\textbf {\bibinfo {volume}
  {107}},\ \bibinfo {pages} {146803} (\bibinfo {year} {2011})}\BibitemShut
  {NoStop}%
\bibitem [{\citenamefont {Wang}\ \emph {et~al.}(2012)\citenamefont {Wang},
  \citenamefont {Yao}, \citenamefont {Gu}, \citenamefont {Gong},\ and\
  \citenamefont {Sheng}}]{Wang2012PhysRevLett.108.126805}%
  \BibitemOpen
  \bibfield  {author} {\bibinfo {author} {\bibfnamefont {Y.-F.}\ \bibnamefont
  {Wang}}, \bibinfo {author} {\bibfnamefont {H.}~\bibnamefont {Yao}}, \bibinfo
  {author} {\bibfnamefont {Z.-C.}\ \bibnamefont {Gu}}, \bibinfo {author}
  {\bibfnamefont {C.-D.}\ \bibnamefont {Gong}},\ and\ \bibinfo {author}
  {\bibfnamefont {D.~N.}\ \bibnamefont {Sheng}},\ }\bibfield  {title} {\bibinfo
  {title} {Non-abelian quantum hall effect in topological flat bands},\ }\href
  {https://doi.org/10.1103/PhysRevLett.108.126805} {\bibfield  {journal}
  {\bibinfo  {journal} {Phys. Rev. Lett.}\ }\textbf {\bibinfo {volume} {108}},\
  \bibinfo {pages} {126805} (\bibinfo {year} {2012})}\BibitemShut {NoStop}%
\bibitem [{\citenamefont {Wu}\ \emph {et~al.}(2012)\citenamefont {Wu},
  \citenamefont {Bernevig},\ and\ \citenamefont
  {Regnault}}]{Wu2012PhysRevB.85.075116}%
  \BibitemOpen
  \bibfield  {author} {\bibinfo {author} {\bibfnamefont {Y.-L.}\ \bibnamefont
  {Wu}}, \bibinfo {author} {\bibfnamefont {B.~A.}\ \bibnamefont {Bernevig}},\
  and\ \bibinfo {author} {\bibfnamefont {N.}~\bibnamefont {Regnault}},\
  }\bibfield  {title} {\bibinfo {title} {Zoology of fractional chern
  insulators},\ }\href {https://doi.org/10.1103/PhysRevB.85.075116} {\bibfield
  {journal} {\bibinfo  {journal} {Phys. Rev. B}\ }\textbf {\bibinfo {volume}
  {85}},\ \bibinfo {pages} {075116} (\bibinfo {year} {2012})}\BibitemShut
  {NoStop}%
\bibitem [{\citenamefont {L\"auchli}\ \emph {et~al.}(2013)\citenamefont
  {L\"auchli}, \citenamefont {Liu}, \citenamefont {Bergholtz},\ and\
  \citenamefont {Moessner}}]{Lauchli2013}%
  \BibitemOpen
  \bibfield  {author} {\bibinfo {author} {\bibfnamefont {A.~M.}\ \bibnamefont
  {L\"auchli}}, \bibinfo {author} {\bibfnamefont {Z.}~\bibnamefont {Liu}},
  \bibinfo {author} {\bibfnamefont {E.~J.}\ \bibnamefont {Bergholtz}},\ and\
  \bibinfo {author} {\bibfnamefont {R.}~\bibnamefont {Moessner}},\ }\bibfield
  {title} {\bibinfo {title} {Hierarchy of fractional chern insulators and
  competing compressible states},\ }\href
  {https://doi.org/10.1103/PhysRevLett.111.126802} {\bibfield  {journal}
  {\bibinfo  {journal} {Phys. Rev. Lett.}\ }\textbf {\bibinfo {volume} {111}},\
  \bibinfo {pages} {126802} (\bibinfo {year} {2013})}\BibitemShut {NoStop}%
\bibitem [{\citenamefont {Liu}\ and\ \citenamefont
  {Bergholtz}(2013)}]{Liu2013PhysRevB.87.035306}%
  \BibitemOpen
  \bibfield  {author} {\bibinfo {author} {\bibfnamefont {Z.}~\bibnamefont
  {Liu}}\ and\ \bibinfo {author} {\bibfnamefont {E.~J.}\ \bibnamefont
  {Bergholtz}},\ }\bibfield  {title} {\bibinfo {title} {From fractional chern
  insulators to abelian and non-abelian fractional quantum hall states:
  Adiabatic continuity and orbital entanglement spectrum},\ }\href
  {https://doi.org/10.1103/PhysRevB.87.035306} {\bibfield  {journal} {\bibinfo
  {journal} {Phys. Rev. B}\ }\textbf {\bibinfo {volume} {87}},\ \bibinfo
  {pages} {035306} (\bibinfo {year} {2013})}\BibitemShut {NoStop}%
\bibitem [{\citenamefont {Sheng}\ \emph {et~al.}(2011)\citenamefont {Sheng},
  \citenamefont {Gu}, \citenamefont {Sun},\ and\ \citenamefont
  {Sheng}}]{Sheng2011NatCommun}%
  \BibitemOpen
  \bibfield  {author} {\bibinfo {author} {\bibfnamefont {D.}~\bibnamefont
  {Sheng}}, \bibinfo {author} {\bibfnamefont {Z.-C.}\ \bibnamefont {Gu}},
  \bibinfo {author} {\bibfnamefont {K.}~\bibnamefont {Sun}},\ and\ \bibinfo
  {author} {\bibfnamefont {L.}~\bibnamefont {Sheng}},\ }\bibfield  {title}
  {\bibinfo {title} {Fractional quantum {{Hall}} effect in the absence of
  {{Landau}} levels},\ }\href {https://doi.org/10.1038/ncomms1380} {\bibfield
  {journal} {\bibinfo  {journal} {Nature Communications}\ }\textbf {\bibinfo
  {volume} {2}},\ \bibinfo {pages} {389} (\bibinfo {year} {2011})}\BibitemShut
  {NoStop}%
\bibitem [{\citenamefont {Kourtis}\ \emph {et~al.}(2014)\citenamefont
  {Kourtis}, \citenamefont {Neupert}, \citenamefont {Chamon},\ and\
  \citenamefont {Mudry}}]{Kourtis2014Phys.Rev.Lett.}%
  \BibitemOpen
  \bibfield  {author} {\bibinfo {author} {\bibfnamefont {S.}~\bibnamefont
  {Kourtis}}, \bibinfo {author} {\bibfnamefont {T.}~\bibnamefont {Neupert}},
  \bibinfo {author} {\bibfnamefont {C.}~\bibnamefont {Chamon}},\ and\ \bibinfo
  {author} {\bibfnamefont {C.}~\bibnamefont {Mudry}},\ }\bibfield  {title}
  {\bibinfo {title} {Fractional {{Chern Insulators}} with {{Strong
  Interactions}} that {{Far Exceed Band Gaps}}},\ }\href
  {https://doi.org/10.1103/PhysRevLett.112.126806} {\bibfield  {journal}
  {\bibinfo  {journal} {Physical Review Letters}\ }\textbf {\bibinfo {volume}
  {112}},\ \bibinfo {pages} {126806} (\bibinfo {year} {2014})}\BibitemShut
  {NoStop}%
\bibitem [{\citenamefont {Grushin}\ \emph {et~al.}(2015)\citenamefont
  {Grushin}, \citenamefont {Motruk}, \citenamefont {Zaletel},\ and\
  \citenamefont {Pollmann}}]{Grushin2015Phys.Rev.B}%
  \BibitemOpen
  \bibfield  {author} {\bibinfo {author} {\bibfnamefont {A.~G.}\ \bibnamefont
  {Grushin}}, \bibinfo {author} {\bibfnamefont {J.}~\bibnamefont {Motruk}},
  \bibinfo {author} {\bibfnamefont {M.~P.}\ \bibnamefont {Zaletel}},\ and\
  \bibinfo {author} {\bibfnamefont {F.}~\bibnamefont {Pollmann}},\ }\bibfield
  {title} {\bibinfo {title} {Characterization and stability of a fermionic
  {$\nu$} = 1 / 3 fractional {{Chern}} insulator},\ }\href
  {https://doi.org/10.1103/PhysRevB.91.035136} {\bibfield  {journal} {\bibinfo
  {journal} {Physical Review B}\ }\textbf {\bibinfo {volume} {91}},\ \bibinfo
  {pages} {035136} (\bibinfo {year} {2015})}\BibitemShut {NoStop}%
\bibitem [{\citenamefont {Morales-Dur\'an}\ \emph {et~al.}(2024)\citenamefont
  {Morales-Dur\'an}, \citenamefont {Wei}, \citenamefont {Shi},\ and\
  \citenamefont {MacDonald}}]{morales2024magic}%
  \BibitemOpen
  \bibfield  {author} {\bibinfo {author} {\bibfnamefont {N.}~\bibnamefont
  {Morales-Dur\'an}}, \bibinfo {author} {\bibfnamefont {N.}~\bibnamefont
  {Wei}}, \bibinfo {author} {\bibfnamefont {J.}~\bibnamefont {Shi}},\ and\
  \bibinfo {author} {\bibfnamefont {A.~H.}\ \bibnamefont {MacDonald}},\
  }\bibfield  {title} {\bibinfo {title} {Magic angles and fractional chern
  insulators in twisted homobilayer transition metal dichalcogenides},\ }\href
  {https://doi.org/10.1103/PhysRevLett.132.096602} {\bibfield  {journal}
  {\bibinfo  {journal} {Phys. Rev. Lett.}\ }\textbf {\bibinfo {volume} {132}},\
  \bibinfo {pages} {096602} (\bibinfo {year} {2024})}\BibitemShut {NoStop}%
\bibitem [{\citenamefont {Yu}\ \emph {et~al.}(2024)\citenamefont {Yu},
  \citenamefont {Herzog-Arbeitman}, \citenamefont {Wang}, \citenamefont
  {Vafek}, \citenamefont {Bernevig},\ and\ \citenamefont
  {Regnault}}]{Yu2024PhysRevB.109.045147}%
  \BibitemOpen
  \bibfield  {author} {\bibinfo {author} {\bibfnamefont {J.}~\bibnamefont
  {Yu}}, \bibinfo {author} {\bibfnamefont {J.}~\bibnamefont
  {Herzog-Arbeitman}}, \bibinfo {author} {\bibfnamefont {M.}~\bibnamefont
  {Wang}}, \bibinfo {author} {\bibfnamefont {O.}~\bibnamefont {Vafek}},
  \bibinfo {author} {\bibfnamefont {B.~A.}\ \bibnamefont {Bernevig}},\ and\
  \bibinfo {author} {\bibfnamefont {N.}~\bibnamefont {Regnault}},\ }\bibfield
  {title} {\bibinfo {title} {Fractional chern insulators versus nonmagnetic
  states in twisted bilayer ${\mathrm{mote}}_{2}$},\ }\href
  {https://doi.org/10.1103/PhysRevB.109.045147} {\bibfield  {journal} {\bibinfo
   {journal} {Phys. Rev. B}\ }\textbf {\bibinfo {volume} {109}},\ \bibinfo
  {pages} {045147} (\bibinfo {year} {2024})}\BibitemShut {NoStop}%
\bibitem [{\citenamefont {Parameswaran}\ \emph {et~al.}(2012)\citenamefont
  {Parameswaran}, \citenamefont {Roy},\ and\ \citenamefont
  {Sondhi}}]{Parameswaran2012PhysRevB.85.241308}%
  \BibitemOpen
  \bibfield  {author} {\bibinfo {author} {\bibfnamefont {S.~A.}\ \bibnamefont
  {Parameswaran}}, \bibinfo {author} {\bibfnamefont {R.}~\bibnamefont {Roy}},\
  and\ \bibinfo {author} {\bibfnamefont {S.~L.}\ \bibnamefont {Sondhi}},\
  }\bibfield  {title} {\bibinfo {title} {Fractional chern insulators and the
  ${W}_{\ensuremath{\infty}}$ algebra},\ }\href
  {https://doi.org/10.1103/PhysRevB.85.241308} {\bibfield  {journal} {\bibinfo
  {journal} {Phys. Rev. B}\ }\textbf {\bibinfo {volume} {85}},\ \bibinfo
  {pages} {241308} (\bibinfo {year} {2012})}\BibitemShut {NoStop}%
\bibitem [{\citenamefont {Kourtis}\ and\ \citenamefont
  {Daghofer}(2014)}]{Kourtis2014Phys.Rev.Lett.a}%
  \BibitemOpen
  \bibfield  {author} {\bibinfo {author} {\bibfnamefont {S.}~\bibnamefont
  {Kourtis}}\ and\ \bibinfo {author} {\bibfnamefont {M.}~\bibnamefont
  {Daghofer}},\ }\bibfield  {title} {\bibinfo {title} {Combined {{Topological}}
  and {{Landau Order}} from {{Strong Correlations}} in {{Chern Bands}}},\
  }\href {https://doi.org/10.1103/PhysRevLett.113.216404} {\bibfield  {journal}
  {\bibinfo  {journal} {Physical Review Letters}\ }\textbf {\bibinfo {volume}
  {113}},\ \bibinfo {pages} {216404} (\bibinfo {year} {2014})}\BibitemShut
  {NoStop}%
\bibitem [{\citenamefont {Claassen}\ \emph {et~al.}(2015)\citenamefont
  {Claassen}, \citenamefont {Lee}, \citenamefont {Thomale}, \citenamefont
  {Qi},\ and\ \citenamefont {Devereaux}}]{Claassen2015PhysRevLett.114.236802}%
  \BibitemOpen
  \bibfield  {author} {\bibinfo {author} {\bibfnamefont {M.}~\bibnamefont
  {Claassen}}, \bibinfo {author} {\bibfnamefont {C.~H.}\ \bibnamefont {Lee}},
  \bibinfo {author} {\bibfnamefont {R.}~\bibnamefont {Thomale}}, \bibinfo
  {author} {\bibfnamefont {X.-L.}\ \bibnamefont {Qi}},\ and\ \bibinfo {author}
  {\bibfnamefont {T.~P.}\ \bibnamefont {Devereaux}},\ }\bibfield  {title}
  {\bibinfo {title} {Position-momentum duality and fractional quantum hall
  effect in chern insulators},\ }\href
  {https://doi.org/10.1103/PhysRevLett.114.236802} {\bibfield  {journal}
  {\bibinfo  {journal} {Phys. Rev. Lett.}\ }\textbf {\bibinfo {volume} {114}},\
  \bibinfo {pages} {236802} (\bibinfo {year} {2015})}\BibitemShut {NoStop}%
\bibitem [{\citenamefont {Wang}\ \emph {et~al.}(2021)\citenamefont {Wang},
  \citenamefont {Cano}, \citenamefont {Millis}, \citenamefont {Liu},\ and\
  \citenamefont {Yang}}]{Wang2021PhysRevLett.127.246403}%
  \BibitemOpen
  \bibfield  {author} {\bibinfo {author} {\bibfnamefont {J.}~\bibnamefont
  {Wang}}, \bibinfo {author} {\bibfnamefont {J.}~\bibnamefont {Cano}}, \bibinfo
  {author} {\bibfnamefont {A.~J.}\ \bibnamefont {Millis}}, \bibinfo {author}
  {\bibfnamefont {Z.}~\bibnamefont {Liu}},\ and\ \bibinfo {author}
  {\bibfnamefont {B.}~\bibnamefont {Yang}},\ }\bibfield  {title} {\bibinfo
  {title} {Exact landau level description of geometry and interaction in a
  flatband},\ }\href {https://doi.org/10.1103/PhysRevLett.127.246403}
  {\bibfield  {journal} {\bibinfo  {journal} {Phys. Rev. Lett.}\ }\textbf
  {\bibinfo {volume} {127}},\ \bibinfo {pages} {246403} (\bibinfo {year}
  {2021})}\BibitemShut {NoStop}%
\bibitem [{\citenamefont {Varjas}\ \emph {et~al.}(2022)\citenamefont {Varjas},
  \citenamefont {Abouelkomsan}, \citenamefont {Yang},\ and\ \citenamefont
  {Bergholtz}}]{Varjas2022SciPostPhys.12.4.118}%
  \BibitemOpen
  \bibfield  {author} {\bibinfo {author} {\bibfnamefont {D.}~\bibnamefont
  {Varjas}}, \bibinfo {author} {\bibfnamefont {A.}~\bibnamefont
  {Abouelkomsan}}, \bibinfo {author} {\bibfnamefont {K.}~\bibnamefont {Yang}},\
  and\ \bibinfo {author} {\bibfnamefont {E.~J.}\ \bibnamefont {Bergholtz}},\
  }\bibfield  {title} {\bibinfo {title} {{Topological lattice models with
  constant Berry curvature}},\ }\href
  {https://doi.org/10.21468/SciPostPhys.12.4.118} {\bibfield  {journal}
  {\bibinfo  {journal} {SciPost Phys.}\ }\textbf {\bibinfo {volume} {12}},\
  \bibinfo {pages} {118} (\bibinfo {year} {2022})}\BibitemShut {NoStop}%
\bibitem [{\citenamefont {Abouelkomsan}\ \emph {et~al.}(2023)\citenamefont
  {Abouelkomsan}, \citenamefont {Yang},\ and\ \citenamefont
  {Bergholtz}}]{Abouelkomsan2023PhysRevResearch.5.L012015}%
  \BibitemOpen
  \bibfield  {author} {\bibinfo {author} {\bibfnamefont {A.}~\bibnamefont
  {Abouelkomsan}}, \bibinfo {author} {\bibfnamefont {K.}~\bibnamefont {Yang}},\
  and\ \bibinfo {author} {\bibfnamefont {E.~J.}\ \bibnamefont {Bergholtz}},\
  }\bibfield  {title} {\bibinfo {title} {Quantum metric induced phases in
  moir\'e materials},\ }\href
  {https://doi.org/10.1103/PhysRevResearch.5.L012015} {\bibfield  {journal}
  {\bibinfo  {journal} {Phys. Rev. Res.}\ }\textbf {\bibinfo {volume} {5}},\
  \bibinfo {pages} {L012015} (\bibinfo {year} {2023})}\BibitemShut {NoStop}%
\bibitem [{\citenamefont {Kourtis}\ \emph {et~al.}(2012)\citenamefont
  {Kourtis}, \citenamefont {Venderbos},\ and\ \citenamefont
  {Daghofer}}]{Kourtis2012Phys.Rev.B}%
  \BibitemOpen
  \bibfield  {author} {\bibinfo {author} {\bibfnamefont {S.}~\bibnamefont
  {Kourtis}}, \bibinfo {author} {\bibfnamefont {J.~W.~F.}\ \bibnamefont
  {Venderbos}},\ and\ \bibinfo {author} {\bibfnamefont {M.}~\bibnamefont
  {Daghofer}},\ }\bibfield  {title} {\bibinfo {title} {Fractional {{Chern}}
  insulator on a triangular lattice of strongly correlated \$\{t\}\_\{2g\}\$
  electrons},\ }\href {https://doi.org/10.1103/PhysRevB.86.235118} {\bibfield
  {journal} {\bibinfo  {journal} {Physical Review B}\ }\textbf {\bibinfo
  {volume} {86}},\ \bibinfo {pages} {235118} (\bibinfo {year}
  {2012})}\BibitemShut {NoStop}%
\bibitem [{\citenamefont {Parker}\ \emph {et~al.}(2021)\citenamefont {Parker},
  \citenamefont {Ledwith}, \citenamefont {Khalaf}, \citenamefont {Soejima},
  \citenamefont {Hauschild}, \citenamefont {Xie}, \citenamefont {Pierce},
  \citenamefont {Zaletel}, \citenamefont {Yacoby},\ and\ \citenamefont
  {Vishwanath}}]{Parker2021a}%
  \BibitemOpen
  \bibfield  {author} {\bibinfo {author} {\bibfnamefont {D.}~\bibnamefont
  {Parker}}, \bibinfo {author} {\bibfnamefont {P.}~\bibnamefont {Ledwith}},
  \bibinfo {author} {\bibfnamefont {E.}~\bibnamefont {Khalaf}}, \bibinfo
  {author} {\bibfnamefont {T.}~\bibnamefont {Soejima}}, \bibinfo {author}
  {\bibfnamefont {J.}~\bibnamefont {Hauschild}}, \bibinfo {author}
  {\bibfnamefont {Y.}~\bibnamefont {Xie}}, \bibinfo {author} {\bibfnamefont
  {A.}~\bibnamefont {Pierce}}, \bibinfo {author} {\bibfnamefont {M.~P.}\
  \bibnamefont {Zaletel}}, \bibinfo {author} {\bibfnamefont {A.}~\bibnamefont
  {Yacoby}},\ and\ \bibinfo {author} {\bibfnamefont {A.}~\bibnamefont
  {Vishwanath}},\ }\href {https://doi.org/10.48550/arXiv.2112.13837} {\bibinfo
  {title} {Field-tuned and zero-field fractional {{Chern}} insulators in magic
  angle graphene}} (\bibinfo {year} {2021}),\ \Eprint
  {https://arxiv.org/abs/2112.13837} {arXiv:2112.13837 [cond-mat]} \BibitemShut
  {NoStop}%
\bibitem [{\citenamefont {Wilhelm}\ \emph {et~al.}(2021)\citenamefont
  {Wilhelm}, \citenamefont {Lang},\ and\ \citenamefont
  {L{\"a}uchli}}]{Wilhelm2021Phys.Rev.Bb}%
  \BibitemOpen
  \bibfield  {author} {\bibinfo {author} {\bibfnamefont {P.}~\bibnamefont
  {Wilhelm}}, \bibinfo {author} {\bibfnamefont {T.~C.}\ \bibnamefont {Lang}},\
  and\ \bibinfo {author} {\bibfnamefont {A.~M.}\ \bibnamefont {L{\"a}uchli}},\
  }\bibfield  {title} {\bibinfo {title} {Interplay of fractional {{Chern}}
  insulator and charge density wave phases in twisted bilayer graphene},\
  }\href {https://doi.org/10.1103/PhysRevB.103.125406} {\bibfield  {journal}
  {\bibinfo  {journal} {Physical Review B}\ }\textbf {\bibinfo {volume}
  {103}},\ \bibinfo {pages} {125406} (\bibinfo {year} {2021})}\BibitemShut
  {NoStop}%
\bibitem [{\citenamefont {Cr\'epel}\ and\ \citenamefont
  {Fu}(2023)}]{Crepel2023PhysRevB.107.L201109}%
  \BibitemOpen
  \bibfield  {author} {\bibinfo {author} {\bibfnamefont {V.}~\bibnamefont
  {Cr\'epel}}\ and\ \bibinfo {author} {\bibfnamefont {L.}~\bibnamefont {Fu}},\
  }\bibfield  {title} {\bibinfo {title} {Anomalous hall metal and fractional
  chern insulator in twisted transition metal dichalcogenides},\ }\href
  {https://doi.org/10.1103/PhysRevB.107.L201109} {\bibfield  {journal}
  {\bibinfo  {journal} {Phys. Rev. B}\ }\textbf {\bibinfo {volume} {107}},\
  \bibinfo {pages} {L201109} (\bibinfo {year} {2023})}\BibitemShut {NoStop}%
\bibitem [{\citenamefont {Lu}\ and\ \citenamefont
  {Santos}(2024)}]{Lu2024PhysRevLett.133.186602}%
  \BibitemOpen
  \bibfield  {author} {\bibinfo {author} {\bibfnamefont {T.}~\bibnamefont
  {Lu}}\ and\ \bibinfo {author} {\bibfnamefont {L.~H.}\ \bibnamefont
  {Santos}},\ }\bibfield  {title} {\bibinfo {title} {Fractional chern
  insulators in twisted bilayer ${\mathrm{mote}}_{2}$: A composite fermion
  perspective},\ }\href {https://doi.org/10.1103/PhysRevLett.133.186602}
  {\bibfield  {journal} {\bibinfo  {journal} {Phys. Rev. Lett.}\ }\textbf
  {\bibinfo {volume} {133}},\ \bibinfo {pages} {186602} (\bibinfo {year}
  {2024})}\BibitemShut {NoStop}%
\bibitem [{\citenamefont {Zeng}\ \emph {et~al.}(2023)\citenamefont {Zeng},
  \citenamefont {Xia}, \citenamefont {Kang}, \citenamefont {Zhu}, \citenamefont
  {Kn{\"u}ppel}, \citenamefont {Vaswani}, \citenamefont {Watanabe},
  \citenamefont {Taniguchi}, \citenamefont {Mak},\ and\ \citenamefont
  {Shan}}]{Zeng2023Nature}%
  \BibitemOpen
  \bibfield  {author} {\bibinfo {author} {\bibfnamefont {Y.}~\bibnamefont
  {Zeng}}, \bibinfo {author} {\bibfnamefont {Z.}~\bibnamefont {Xia}}, \bibinfo
  {author} {\bibfnamefont {K.}~\bibnamefont {Kang}}, \bibinfo {author}
  {\bibfnamefont {J.}~\bibnamefont {Zhu}}, \bibinfo {author} {\bibfnamefont
  {P.}~\bibnamefont {Kn{\"u}ppel}}, \bibinfo {author} {\bibfnamefont
  {C.}~\bibnamefont {Vaswani}}, \bibinfo {author} {\bibfnamefont
  {K.}~\bibnamefont {Watanabe}}, \bibinfo {author} {\bibfnamefont
  {T.}~\bibnamefont {Taniguchi}}, \bibinfo {author} {\bibfnamefont {K.~F.}\
  \bibnamefont {Mak}},\ and\ \bibinfo {author} {\bibfnamefont {J.}~\bibnamefont
  {Shan}},\ }\bibfield  {title} {\bibinfo {title} {Thermodynamic evidence of
  fractional {{Chern}} insulator in moir{\'e} {{MoTe2}}},\ }\href
  {https://doi.org/10.1038/s41586-023-06452-3} {\bibfield  {journal} {\bibinfo
  {journal} {Nature}\ }\textbf {\bibinfo {volume} {622}},\ \bibinfo {pages}
  {69} (\bibinfo {year} {2023})}\BibitemShut {NoStop}%
\bibitem [{\citenamefont {Cai}\ \emph {et~al.}(2023)\citenamefont {Cai},
  \citenamefont {Anderson}, \citenamefont {Wang}, \citenamefont {Zhang},
  \citenamefont {Liu}, \citenamefont {Holtzmann}, \citenamefont {Zhang},
  \citenamefont {Fan}, \citenamefont {Taniguchi}, \citenamefont {Watanabe}
  \emph {et~al.}}]{cai2023signatures}%
  \BibitemOpen
  \bibfield  {author} {\bibinfo {author} {\bibfnamefont {J.}~\bibnamefont
  {Cai}}, \bibinfo {author} {\bibfnamefont {E.}~\bibnamefont {Anderson}},
  \bibinfo {author} {\bibfnamefont {C.}~\bibnamefont {Wang}}, \bibinfo {author}
  {\bibfnamefont {X.}~\bibnamefont {Zhang}}, \bibinfo {author} {\bibfnamefont
  {X.}~\bibnamefont {Liu}}, \bibinfo {author} {\bibfnamefont {W.}~\bibnamefont
  {Holtzmann}}, \bibinfo {author} {\bibfnamefont {Y.}~\bibnamefont {Zhang}},
  \bibinfo {author} {\bibfnamefont {F.}~\bibnamefont {Fan}}, \bibinfo {author}
  {\bibfnamefont {T.}~\bibnamefont {Taniguchi}}, \bibinfo {author}
  {\bibfnamefont {K.}~\bibnamefont {Watanabe}}, \emph {et~al.},\ }\bibfield
  {title} {\bibinfo {title} {Signatures of fractional quantum anomalous hall
  states in twisted mote2},\ }\href
  {https://doi.org/10.1038/s41586-023-06289-w} {\bibfield  {journal} {\bibinfo
  {journal} {Nature}\ }\textbf {\bibinfo {volume} {622}},\ \bibinfo {pages}
  {63} (\bibinfo {year} {2023})}\BibitemShut {NoStop}%
\bibitem [{\citenamefont {Ji}\ \emph {et~al.}(2024)\citenamefont {Ji},
  \citenamefont {Park}, \citenamefont {Barber}, \citenamefont {Hu},
  \citenamefont {Watanabe}, \citenamefont {Taniguchi}, \citenamefont {Chu},
  \citenamefont {Xu},\ and\ \citenamefont {Shen}}]{ji2024local}%
  \BibitemOpen
  \bibfield  {author} {\bibinfo {author} {\bibfnamefont {Z.}~\bibnamefont
  {Ji}}, \bibinfo {author} {\bibfnamefont {H.}~\bibnamefont {Park}}, \bibinfo
  {author} {\bibfnamefont {M.~E.}\ \bibnamefont {Barber}}, \bibinfo {author}
  {\bibfnamefont {C.}~\bibnamefont {Hu}}, \bibinfo {author} {\bibfnamefont
  {K.}~\bibnamefont {Watanabe}}, \bibinfo {author} {\bibfnamefont
  {T.}~\bibnamefont {Taniguchi}}, \bibinfo {author} {\bibfnamefont {J.-H.}\
  \bibnamefont {Chu}}, \bibinfo {author} {\bibfnamefont {X.}~\bibnamefont
  {Xu}},\ and\ \bibinfo {author} {\bibfnamefont {Z.-X.}\ \bibnamefont {Shen}},\
  }\bibfield  {title} {\bibinfo {title} {Local probe of bulk and edge states in
  a fractional chern insulator},\ }\href
  {https://doi.org/10.1038/s41586-024-08092-7} {\bibfield  {journal} {\bibinfo
  {journal} {Nature}\ }\textbf {\bibinfo {volume} {635}},\ \bibinfo {pages}
  {578} (\bibinfo {year} {2024})}\BibitemShut {NoStop}%
\bibitem [{\citenamefont {Xie}\ \emph {et~al.}(2021)\citenamefont {Xie},
  \citenamefont {Pierce}, \citenamefont {Park}, \citenamefont {Parker},
  \citenamefont {Khalaf}, \citenamefont {Ledwith}, \citenamefont {Cao},
  \citenamefont {Lee}, \citenamefont {Chen}, \citenamefont {Forrester},
  \citenamefont {Watanabe}, \citenamefont {Taniguchi}, \citenamefont
  {Vishwanath}, \citenamefont {{Jarillo-Herrero}},\ and\ \citenamefont
  {Yacoby}}]{Xie2021Naturea}%
  \BibitemOpen
  \bibfield  {author} {\bibinfo {author} {\bibfnamefont {Y.}~\bibnamefont
  {Xie}}, \bibinfo {author} {\bibfnamefont {A.~T.}\ \bibnamefont {Pierce}},
  \bibinfo {author} {\bibfnamefont {J.~M.}\ \bibnamefont {Park}}, \bibinfo
  {author} {\bibfnamefont {D.~E.}\ \bibnamefont {Parker}}, \bibinfo {author}
  {\bibfnamefont {E.}~\bibnamefont {Khalaf}}, \bibinfo {author} {\bibfnamefont
  {P.}~\bibnamefont {Ledwith}}, \bibinfo {author} {\bibfnamefont
  {Y.}~\bibnamefont {Cao}}, \bibinfo {author} {\bibfnamefont {S.~H.}\
  \bibnamefont {Lee}}, \bibinfo {author} {\bibfnamefont {S.}~\bibnamefont
  {Chen}}, \bibinfo {author} {\bibfnamefont {P.~R.}\ \bibnamefont {Forrester}},
  \bibinfo {author} {\bibfnamefont {K.}~\bibnamefont {Watanabe}}, \bibinfo
  {author} {\bibfnamefont {T.}~\bibnamefont {Taniguchi}}, \bibinfo {author}
  {\bibfnamefont {A.}~\bibnamefont {Vishwanath}}, \bibinfo {author}
  {\bibfnamefont {P.}~\bibnamefont {{Jarillo-Herrero}}},\ and\ \bibinfo
  {author} {\bibfnamefont {A.}~\bibnamefont {Yacoby}},\ }\bibfield  {title}
  {\bibinfo {title} {Fractional {{Chern}} insulators in magic-angle twisted
  bilayer graphene},\ }\href {https://doi.org/10.1038/s41586-021-04002-3}
  {\bibfield  {journal} {\bibinfo  {journal} {Nature}\ }\textbf {\bibinfo
  {volume} {600}},\ \bibinfo {pages} {439} (\bibinfo {year}
  {2021})}\BibitemShut {NoStop}%
\bibitem [{\citenamefont {Aronson}\ \emph {et~al.}(2024)\citenamefont
  {Aronson}, \citenamefont {Han}, \citenamefont {Lu}, \citenamefont {Yao},
  \citenamefont {Watanabe}, \citenamefont {Taniguchi}, \citenamefont {Ju},\
  and\ \citenamefont {Ashoori}}]{Aronson2024}%
  \BibitemOpen
  \bibfield  {author} {\bibinfo {author} {\bibfnamefont {S.~H.}\ \bibnamefont
  {Aronson}}, \bibinfo {author} {\bibfnamefont {T.}~\bibnamefont {Han}},
  \bibinfo {author} {\bibfnamefont {Z.}~\bibnamefont {Lu}}, \bibinfo {author}
  {\bibfnamefont {Y.}~\bibnamefont {Yao}}, \bibinfo {author} {\bibfnamefont
  {K.}~\bibnamefont {Watanabe}}, \bibinfo {author} {\bibfnamefont
  {T.}~\bibnamefont {Taniguchi}}, \bibinfo {author} {\bibfnamefont
  {L.}~\bibnamefont {Ju}},\ and\ \bibinfo {author} {\bibfnamefont {R.~C.}\
  \bibnamefont {Ashoori}},\ }\href {https://doi.org/10.48550/arXiv.2408.11220}
  {\bibinfo {title} {Displacement field-controlled fractional {{Chern}}
  insulators and charge density waves in a graphene/{{hBN}} moir{\'e}
  superlattice}} (\bibinfo {year} {2024}),\ \Eprint
  {https://arxiv.org/abs/2408.11220} {arXiv:2408.11220} \BibitemShut {NoStop}%
\bibitem [{\citenamefont {Zhang}\ \emph {et~al.}(2024)\citenamefont {Zhang},
  \citenamefont {Mao}, \citenamefont {Kim},\ and\ \citenamefont
  {Moessner}}]{zhang2024}%
  \BibitemOpen
  \bibfield  {author} {\bibinfo {author} {\bibfnamefont {K.}~\bibnamefont
  {Zhang}}, \bibinfo {author} {\bibfnamefont {D.}~\bibnamefont {Mao}}, \bibinfo
  {author} {\bibfnamefont {E.-A.}\ \bibnamefont {Kim}},\ and\ \bibinfo {author}
  {\bibfnamefont {R.}~\bibnamefont {Moessner}},\ }\href
  {https://arxiv.org/abs/2410.00092} {\bibinfo {title} {Bionic
  fractionalization in the trimer model of twisted bilayer graphene}} (\bibinfo
  {year} {2024}),\ \Eprint {https://arxiv.org/abs/2410.00092} {arXiv:2410.00092
  [cond-mat.str-el]} \BibitemShut {NoStop}%
\bibitem [{\citenamefont {Po}\ \emph {et~al.}(2018)\citenamefont {Po},
  \citenamefont {Zou}, \citenamefont {Vishwanath},\ and\ \citenamefont
  {Senthil}}]{po2018origin}%
  \BibitemOpen
  \bibfield  {author} {\bibinfo {author} {\bibfnamefont {H.~C.}\ \bibnamefont
  {Po}}, \bibinfo {author} {\bibfnamefont {L.}~\bibnamefont {Zou}}, \bibinfo
  {author} {\bibfnamefont {A.}~\bibnamefont {Vishwanath}},\ and\ \bibinfo
  {author} {\bibfnamefont {T.}~\bibnamefont {Senthil}},\ }\bibfield  {title}
  {\bibinfo {title} {Origin of mott insulating behavior and superconductivity
  in twisted bilayer graphene},\ }\href
  {https://doi.org/10.1103/PhysRevX.8.031089} {\bibfield  {journal} {\bibinfo
  {journal} {Physical Review X}\ }\textbf {\bibinfo {volume} {8}},\ \bibinfo
  {pages} {031089} (\bibinfo {year} {2018})}\BibitemShut {NoStop}%
\bibitem [{\citenamefont {Kang}\ and\ \citenamefont
  {Vafek}(2019)}]{Kang2019PhysRevLett.122.246401}%
  \BibitemOpen
  \bibfield  {author} {\bibinfo {author} {\bibfnamefont {J.}~\bibnamefont
  {Kang}}\ and\ \bibinfo {author} {\bibfnamefont {O.}~\bibnamefont {Vafek}},\
  }\bibfield  {title} {\bibinfo {title} {Strong coupling phases of partially
  filled twisted bilayer graphene narrow bands},\ }\href
  {https://doi.org/10.1103/PhysRevLett.122.246401} {\bibfield  {journal}
  {\bibinfo  {journal} {Phys. Rev. Lett.}\ }\textbf {\bibinfo {volume} {122}},\
  \bibinfo {pages} {246401} (\bibinfo {year} {2019})}\BibitemShut {NoStop}%
\bibitem [{\citenamefont {Zhang}\ \emph {et~al.}(2019)\citenamefont {Zhang},
  \citenamefont {Mao},\ and\ \citenamefont {Senthil}}]{zhang2019twisted}%
  \BibitemOpen
  \bibfield  {author} {\bibinfo {author} {\bibfnamefont {Y.-H.}\ \bibnamefont
  {Zhang}}, \bibinfo {author} {\bibfnamefont {D.}~\bibnamefont {Mao}},\ and\
  \bibinfo {author} {\bibfnamefont {T.}~\bibnamefont {Senthil}},\ }\bibfield
  {title} {\bibinfo {title} {Twisted bilayer graphene aligned with hexagonal
  boron nitride: Anomalous hall effect and a lattice model},\ }\href
  {https://doi.org/10.1103/PhysRevResearch.1.033126} {\bibfield  {journal}
  {\bibinfo  {journal} {Physical Review Research}\ }\textbf {\bibinfo {volume}
  {1}},\ \bibinfo {pages} {033126} (\bibinfo {year} {2019})}\BibitemShut
  {NoStop}%
\bibitem [{\citenamefont {Astrakhantsev}\ \emph {et~al.}(2023)\citenamefont
  {Astrakhantsev}, \citenamefont {Wagner}, \citenamefont {Westerhout},
  \citenamefont {Neupert},\ and\ \citenamefont {Fischer}}]{Astrakhantsev2023}%
  \BibitemOpen
  \bibfield  {author} {\bibinfo {author} {\bibfnamefont {N.}~\bibnamefont
  {Astrakhantsev}}, \bibinfo {author} {\bibfnamefont {G.}~\bibnamefont
  {Wagner}}, \bibinfo {author} {\bibfnamefont {T.}~\bibnamefont {Westerhout}},
  \bibinfo {author} {\bibfnamefont {T.}~\bibnamefont {Neupert}},\ and\ \bibinfo
  {author} {\bibfnamefont {M.~H.}\ \bibnamefont {Fischer}},\ }\bibfield
  {title} {\bibinfo {title} {Understanding symmetry breaking in twisted bilayer
  graphene from cluster constraints},\ }\href
  {https://doi.org/10.1103/PhysRevResearch.5.043214} {\bibfield  {journal}
  {\bibinfo  {journal} {Phys. Rev. Res.}\ }\textbf {\bibinfo {volume} {5}},\
  \bibinfo {pages} {043214} (\bibinfo {year} {2023})}\BibitemShut {NoStop}%
\bibitem [{\citenamefont {Verberkmoes}\ and\ \citenamefont
  {Nienhuis}(1999)}]{verberkmoes1999triangular}%
  \BibitemOpen
  \bibfield  {author} {\bibinfo {author} {\bibfnamefont {A.}~\bibnamefont
  {Verberkmoes}}\ and\ \bibinfo {author} {\bibfnamefont {B.}~\bibnamefont
  {Nienhuis}},\ }\bibfield  {title} {\bibinfo {title} {Triangular trimers on
  the triangular lattice: An exact solution},\ }\href
  {https://doi.org/10.1103/PhysRevLett.83.3986} {\bibfield  {journal} {\bibinfo
   {journal} {Physical review letters}\ }\textbf {\bibinfo {volume} {83}},\
  \bibinfo {pages} {3986} (\bibinfo {year} {1999})}\BibitemShut {NoStop}%
\bibitem [{\citenamefont {Zhang}\ \emph {et~al.}(2022)\citenamefont {Zhang},
  \citenamefont {Zhang}, \citenamefont {Fu},\ and\ \citenamefont
  {Kim}}]{zhang2022fractional}%
  \BibitemOpen
  \bibfield  {author} {\bibinfo {author} {\bibfnamefont {K.}~\bibnamefont
  {Zhang}}, \bibinfo {author} {\bibfnamefont {Y.}~\bibnamefont {Zhang}},
  \bibinfo {author} {\bibfnamefont {L.}~\bibnamefont {Fu}},\ and\ \bibinfo
  {author} {\bibfnamefont {E.-A.}\ \bibnamefont {Kim}},\ }\bibfield  {title}
  {\bibinfo {title} {Fractional correlated insulating states at one-third
  filled magic angle twisted bilayer graphene},\ }\href
  {https://doi.org/10.1038/s42005-022-01027-6} {\bibfield  {journal} {\bibinfo
  {journal} {Communications Physics}\ }\textbf {\bibinfo {volume} {5}},\
  \bibinfo {pages} {250} (\bibinfo {year} {2022})}\BibitemShut {NoStop}%
\bibitem [{\citenamefont {Rokhsar}\ and\ \citenamefont {Kivelson}(1988)}]{RK}%
  \BibitemOpen
  \bibfield  {author} {\bibinfo {author} {\bibfnamefont {D.~S.}\ \bibnamefont
  {Rokhsar}}\ and\ \bibinfo {author} {\bibfnamefont {S.~A.}\ \bibnamefont
  {Kivelson}},\ }\bibfield  {title} {\bibinfo {title} {Superconductivity and
  the quantum hard-core dimer gas},\ }\href
  {https://doi.org/10.1103/PhysRevLett.61.2376} {\bibfield  {journal} {\bibinfo
   {journal} {Phys. Rev. Lett.}\ }\textbf {\bibinfo {volume} {61}},\ \bibinfo
  {pages} {2376} (\bibinfo {year} {1988})}\BibitemShut {NoStop}%
\bibitem [{\citenamefont {Moessner}\ \emph {et~al.}(2001)\citenamefont
  {Moessner}, \citenamefont {Sondhi},\ and\ \citenamefont
  {Fradkin}}]{moessner2001short}%
  \BibitemOpen
  \bibfield  {author} {\bibinfo {author} {\bibfnamefont {R.}~\bibnamefont
  {Moessner}}, \bibinfo {author} {\bibfnamefont {S.~L.}\ \bibnamefont
  {Sondhi}},\ and\ \bibinfo {author} {\bibfnamefont {E.}~\bibnamefont
  {Fradkin}},\ }\bibfield  {title} {\bibinfo {title} {Short-ranged resonating
  valence bond physics, quantum dimer models, and ising gauge theories},\
  }\href {https://doi.org/10.1103/PhysRevB.65.024504} {\bibfield  {journal}
  {\bibinfo  {journal} {Physical Review B}\ }\textbf {\bibinfo {volume} {65}},\
  \bibinfo {pages} {024504} (\bibinfo {year} {2001})}\BibitemShut {NoStop}%
\bibitem [{\citenamefont {Repellin}\ \emph {et~al.}(2014)\citenamefont
  {Repellin}, \citenamefont {Bernevig},\ and\ \citenamefont
  {Regnault}}]{Repellin2014}%
  \BibitemOpen
  \bibfield  {author} {\bibinfo {author} {\bibfnamefont {C.}~\bibnamefont
  {Repellin}}, \bibinfo {author} {\bibfnamefont {B.~A.}\ \bibnamefont
  {Bernevig}},\ and\ \bibinfo {author} {\bibfnamefont {N.}~\bibnamefont
  {Regnault}},\ }\bibfield  {title} {\bibinfo {title} {{${\mathbb{Z}}_{2}$}
  fractional topological insulators in two dimensions},\ }\href
  {https://doi.org/10.1103/PhysRevB.90.245401} {\bibfield  {journal} {\bibinfo
  {journal} {Phys. Rev. B}\ }\textbf {\bibinfo {volume} {90}},\ \bibinfo
  {pages} {245401} (\bibinfo {year} {2014})}\BibitemShut {NoStop}%
\bibitem [{si()}]{si}%
  \BibitemOpen
  \href@noop {} {}\bibinfo {note} {See supplementary material for additional
  details}\BibitemShut {NoStop}%
\bibitem [{\citenamefont {Mao}\ \emph {et~al.}(2023)\citenamefont {Mao},
  \citenamefont {Zhang},\ and\ \citenamefont {Kim}}]{Mao2023}%
  \BibitemOpen
  \bibfield  {author} {\bibinfo {author} {\bibfnamefont {D.}~\bibnamefont
  {Mao}}, \bibinfo {author} {\bibfnamefont {K.}~\bibnamefont {Zhang}},\ and\
  \bibinfo {author} {\bibfnamefont {E.-A.}\ \bibnamefont {Kim}},\ }\bibfield
  {title} {\bibinfo {title} {Fractionalization in fractional correlated
  insulating states at $n\ifmmode\pm\else\textpm\fi{}1/3$ filled twisted
  bilayer graphene},\ }\href {https://doi.org/10.1103/PhysRevLett.131.106801}
  {\bibfield  {journal} {\bibinfo  {journal} {Phys. Rev. Lett.}\ }\textbf
  {\bibinfo {volume} {131}},\ \bibinfo {pages} {106801} (\bibinfo {year}
  {2023})}\BibitemShut {NoStop}%
\bibitem [{\citenamefont {Bravyi}\ \emph {et~al.}(2011)\citenamefont {Bravyi},
  \citenamefont {DiVincenzo},\ and\ \citenamefont {Loss}}]{Bravyi2011}%
  \BibitemOpen
  \bibfield  {author} {\bibinfo {author} {\bibfnamefont {S.}~\bibnamefont
  {Bravyi}}, \bibinfo {author} {\bibfnamefont {D.~P.}\ \bibnamefont
  {DiVincenzo}},\ and\ \bibinfo {author} {\bibfnamefont {D.}~\bibnamefont
  {Loss}},\ }\bibfield  {title} {\bibinfo {title} {Schrieffer--wolff
  transformation for quantum many-body systems},\ }\href
  {https://doi.org/https://doi.org/10.1016/j.aop.2011.06.004} {\bibfield
  {journal} {\bibinfo  {journal} {Annals of physics}\ }\textbf {\bibinfo
  {volume} {326}},\ \bibinfo {pages} {2793} (\bibinfo {year}
  {2011})}\BibitemShut {NoStop}%
\bibitem [{\citenamefont {Slagle}\ and\ \citenamefont
  {Kim}(2017)}]{Slagle2017}%
  \BibitemOpen
  \bibfield  {author} {\bibinfo {author} {\bibfnamefont {K.}~\bibnamefont
  {Slagle}}\ and\ \bibinfo {author} {\bibfnamefont {Y.~B.}\ \bibnamefont
  {Kim}},\ }\bibfield  {title} {\bibinfo {title} {Fracton topological order
  from nearest-neighbor two-spin interactions and dualities},\ }\href
  {https://doi.org/10.1103/PhysRevB.96.165106} {\bibfield  {journal} {\bibinfo
  {journal} {Phys. Rev. B}\ }\textbf {\bibinfo {volume} {96}},\ \bibinfo
  {pages} {165106} (\bibinfo {year} {2017})}\BibitemShut {NoStop}%
\bibitem [{\citenamefont {Musser}\ \emph {et~al.}(2024)\citenamefont {Musser},
  \citenamefont {Cheng},\ and\ \citenamefont
  {Senthil}}]{musser2024fractionalizationalternatechargeordering}%
  \BibitemOpen
  \bibfield  {author} {\bibinfo {author} {\bibfnamefont {S.}~\bibnamefont
  {Musser}}, \bibinfo {author} {\bibfnamefont {M.}~\bibnamefont {Cheng}},\ and\
  \bibinfo {author} {\bibfnamefont {T.}~\bibnamefont {Senthil}},\ }\href
  {https://arxiv.org/abs/2408.03984} {\bibinfo {title} {Fractionalization as an
  alternate to charge ordering in electronic insulators}} (\bibinfo {year}
  {2024}),\ \Eprint {https://arxiv.org/abs/2408.03984} {arXiv:2408.03984
  [cond-mat.str-el]} \BibitemShut {NoStop}%
\bibitem [{\citenamefont {Fukui}\ \emph {et~al.}(2005)\citenamefont {Fukui},
  \citenamefont {Hatsugai},\ and\ \citenamefont {Suzuki}}]{fukui2005chern}%
  \BibitemOpen
  \bibfield  {author} {\bibinfo {author} {\bibfnamefont {T.}~\bibnamefont
  {Fukui}}, \bibinfo {author} {\bibfnamefont {Y.}~\bibnamefont {Hatsugai}},\
  and\ \bibinfo {author} {\bibfnamefont {H.}~\bibnamefont {Suzuki}},\
  }\bibfield  {title} {\bibinfo {title} {Chern numbers in discretized brillouin
  zone: efficient method of computing (spin) hall conductances},\ }\href
  {https://doi.org/10.1143/JPSJ.74.1674} {\bibfield  {journal} {\bibinfo
  {journal} {Journal of the Physical Society of Japan}\ }\textbf {\bibinfo
  {volume} {74}},\ \bibinfo {pages} {1674} (\bibinfo {year}
  {2005})}\BibitemShut {NoStop}%
\end{thebibliography}%

\clearpage

\renewcommand{\thesection}{\Alph{section}}
\setcounter{figure}{0}
\setcounter{equation}{0}
\renewcommand\thefigure{A\arabic{figure}}
\renewcommand\theequation{A\arabic{equation}}
\setcounter{table}{0}
\renewcommand\thetable{A\arabic{table}}

\begin{widetext}
\section*{Appendix A: Tilted boundary conditions}
\label{app:Tilted_BC}
The tilted boundary conditions (TBC) generalize the usual periodic boundary conditions (PBC) and are used to generate more isotropic clusters given a fixed number of sites. In standard PBC, the periodic vectors $\mathbf{T}_1$ and $\mathbf{T}_2$ for an $N_1\times N_2$ cluster are fixed to align with the primitive lattice vectors:
\begin{equation}
\mathbf{T}_1=N_1 \mathbf{a}_1, \quad \mathbf{T}_2 = N_2 \mathbf{a}_2.
\end{equation}
The symmetric lattice translation generators $\mathbf{t}_1$ and $\mathbf{t_2}$, which are used to implement the lattice translation symmetry, satisfy the following conditions:
\begin{enumerate}
\item They map unit cells to unit cells, i.e., $\mathbf{t}_1$ and $\mathbf{t}_2$ can be expressed as integer linear combinations of $\mathbf{a}_1$ and $\mathbf{a}_2$.
\item They span the area of that of a unit cell: $|\mathbf{t}_1\times \mathbf{t}_2| = |\mathbf{a}_1\times \mathbf{a}_2 |$.
\item  $N_1 \mathbf{t}_1$ and $N_2 \mathbf{t}_2$ map unit cells back to themselves under the periodic boundary conditions specified by $\mathbf{T}_1$ and $\mathbf{T}_2$.
\end{enumerate}
Under PBC, the natural choice for the generators is $\mathbf{t}_1 = \mathbf{a}_1, \mathbf{t}_2 = \mathbf{a}_2$. TBC allow the periodic vectors $\mathbf{T}_1$ and $\mathbf{T}_2$ to be integer linear combinations of $\mathbf{a}_1$ and $\mathbf{a}_2$, with the constraint that the corresponding $\mathbf{t}_1$ and $\mathbf{t}_2$ exist.
In practice, for simplicity, we search for clusters with the symmetric generators having the form $\mathbf{t}_1 = \mathbf{a}_1$ and $\mathbf{t}_2 = \mathbf{a}_2 + \alpha \mathbf{a}_1, \alpha \in \mathbb{Z}$. In Table~\ref{tab:TBC_table} we list the data of the clusters used in this work.

\begin{table}[h!]
    \centering
    \begin{tabular}{|c|c|c|c|c|c|}
        \hline
        \(N_1 \times N_2\) & \(\mathbf{T}_1\) & \(\mathbf{T}_2\) & \(\alpha\) & deg & asp. ratio\\
        \hline
        9 $\times$ 2 & \(\mathbf{a}_1 + 4 \mathbf{a}_2\) & \(-4 \mathbf{a}_1 + 2\mathbf{a}_1\) & -2 & 48 & 1.35 \\
        \hline
        21 $\times$ 1 & \(4 \mathbf{a}_1 + \mathbf{a}_2\) & \(-5 \mathbf{a}_1 + 4\mathbf{a}_1\) & 4 & 48 & 1.15 \\
        \hline
        24 $\times$ 1 & \(2 \mathbf{a}_1 + 2\mathbf{a}_2\) & \(7 \mathbf{a}_1  -5 \mathbf{a}_2\) & 13 & 96 &0.58\\
        \hline
        24 $\times$ 1 & \(4 \mathbf{a}_1 + \mathbf{a}_2\) & \(4 \mathbf{a}_1  -5 \mathbf{a}_2\) & 4 & 24 &1.01\\
        \hline
        12 $\times$ 2 & \(6 \mathbf{a}_2\) & \(4 \mathbf{a}_1 - 2 \mathbf{a}_2\) & 10 & 96 & 1.73 \\
        \hline
    \end{tabular}
    \caption{
    Summary of the clusters used in this work. The vectors $\mathbf{T}_1$ and $\mathbf{T}_2$ are the periodic vectors specifying the periodic boundary conditions. The integer $\alpha$ determines the symmetric translation generators $\mathbf{t}_1 = \mathbf{a}_1$ and $\mathbf{t}_2 = \mathbf{a}_2 + \alpha \mathbf{a}_1$. The column labeled ``deg" represents the ground state degeneracy of the classical cluster-charging Hamiltonian on the finite clusters. 
    The last column lists the aspect ratios of the clusters.
    }
    \label{tab:TBC_table}
\end{table}

\setcounter{figure}{0}
\setcounter{equation}{0}
\renewcommand\thefigure{B\arabic{figure}}
\renewcommand\theequation{B\arabic{equation}}
\section*{Appendix B: Calculation of many-body Chern number}
\label{app:chern}
This section provides details on calculating the Chern number $\mathcal{C}$ for a $m$-fold degenerate ground state manifold in ED.     
To calculate $\mathcal{C}$, we introduce flux $\boldsymbol{\Phi} = (\Phi_x, \Phi_y) \in [0, 2\pi]^{\times 2}$ along the two periodic directions of the torus, which modifies the phases of the hopping amplitudes. 
Specifically, for a (possibly tilted) $N_1\times N_2$ torus,
we have two symmetric translation vectors $\mathbf{t_1}$ and $\mathbf{t_2}$ satisfying that (a) 
they span the area of a unit cell $|\mathbf{t_1}\times \mathbf{t_2}| = |\mathbf{a_1}\times \mathbf{a_2} |$ ($\mathbf{a_1}$ and $\mathbf{a_2}$ are primitive lattice vectors)
and (b) $\mathbf{T_1} = N_1 \mathbf{t_1}$ and $\mathbf{T}_2 = N_2 \mathbf{t_2}$ map unit cells back to themselves under the periodic boundary conditions. For untilted clusters, the symmetric translation vectors equal to the primitive lattice vectors $\mathbf{t_1} = \mathbf{a_1}$ and $\mathbf{t_2} = \mathbf{a_2}$.
The phase acquired for a generic hopping term through flux insertion is $n_1 \frac{\Phi_x}{N_1} + n_2 \frac{\Phi_y}{N_2}$, where $\mathbf{r} = n_1 \mathbf{t}_1 + n_2 \mathbf{t}_2$ is the translation vector associated to the hopping.
Because the Hamiltonian remains invariant upon an insertion of a flux quantum
\begin{equation}
    H(\Phi_x, \Phi_y) \sim H(\Phi_x + 2\pi, \Phi_y) \sim H(\Phi_x, \Phi_y + 2\pi),
\end{equation}
where $\sim$ means that the two Hamiltonians are equivalent up to a gauge transformation, the space of inserted flux forms a torus $T^2_{\Phi}$.
Hence, we can define a non-Abelian Berry curvature $\mathcal{F}(\mathbf \Phi)$ for the $m$-fold degenerate ground states on $T^2_{\Phi}$:
\begin{equation}\label{eq:Berry_curv}
       \begin{aligned}
       &\mathcal{F}(\boldsymbol{\Phi})=\partial_{\Phi_x}\mathcal{A}_y(\boldsymbol{\Phi})-
        \partial_{\Phi_y}\mathcal{A}_x(\boldsymbol{\Phi})-i[\mathcal{A}_x(\boldsymbol{\Phi}),
        \mathcal{A}_y(\boldsymbol{\Phi})],\\&\mathcal{A}_j(\boldsymbol{\Phi})=-
        i\boldsymbol{\Psi}_\Phi^\dagger\partial_{\Phi_j}\boldsymbol{\Psi}_\Phi,\quad j=x,y,
       \end{aligned}
\end{equation}
where $\boldsymbol{\Psi}_{\Phi}$ is an orthonormal basis for the ground state space at flux $\mathbf{\Phi}$
\begin{equation}
\boldsymbol{\Psi}_{\Phi} = (|\psi_1 (\boldsymbol \Phi)\rangle, \cdots, |\psi_m (\boldsymbol \Phi)\rangle).
\end{equation}
The Chern number $\mathcal{C}$ is obtained by integrating $\mathcal{F}$ over the flux torus
    \begin{equation}\label{eq:chern}
        \mathcal{C}=\frac1{2\pi}\int_0^{2\pi}\int_0^{2\pi}\mathrm{Tr}[
        \mathcal{F}(\boldsymbol{\Phi})]\mathrm{d}^2\boldsymbol{\Phi},
    \end{equation}
which is quantized to integers. Eq.~\eqref{eq:chern} is evaluated numerically by discretizing $T_{\Phi}^2$ \cite{fukui2005chern} and computing $\mathrm{Tr}[\mathcal{F}(\boldsymbol{\Phi})]$ grid by grid.
 The Hall conductivity is related to $\mathcal{C}$ by $\sigma_{xy} = \frac{\mathcal{C}}{m} \frac{e^2}{h}$.
 Figure~\ref{fig:TrF_9x2} plots $\text{Tr}(\mathcal{F})$ for the gWC phase ($t_1/V_1=0.07$) and FCI phase ($t_1/V_1=0.20$) of the $9\times 2$ cluster defined in the main text.
 
\begin{figure}[htb]
    \centering
    \includegraphics[width=.35\textwidth]{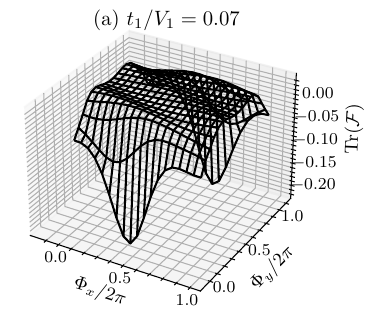}
    \hspace{0.08\textwidth}
    \includegraphics[width=.35\textwidth]{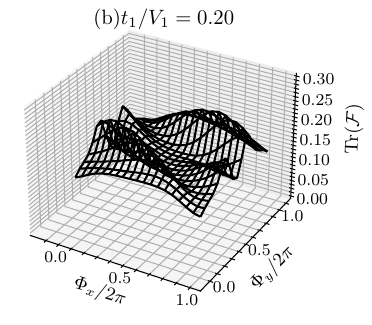}
    \caption{Berry curvature $\text{Tr}(\mathcal{F})$ of the three-fold degenerate ground states on the $9\times 2$ cluster at (a) $t_1/V_1 = 0.07$ and (b) $t_1/V_1 = 0.20$. The data is obtained on a $20\times 20$ grids.
    Integration of $\text{Tr}(\mathcal{F})$ yields $\mathcal{C}=0$ for (a) and $\mathcal{C}=1$ for (b).}
    \label{fig:TrF_9x2}
\end{figure} 

\setcounter{figure}{0}
\setcounter{equation}{0}
\renewcommand\thefigure{C\arabic{figure}}
\renewcommand\theequation{C\arabic{equation}}

\section*{Appendix C: Perturbative treatment}
\label{app:perturbation}
In this section we review the Schrieffer-Wolff (SW) transformation \cite{Bravyi2011, Slagle2017} used for degenerate perturbation theory and apply it to the model introduced in the main text. The method starts with a Hamiltonian $H = H_0 + \lambda H_1$, which is divided into unperturbed part $H_0$ and the perturbation $\lambda H_1$.
We denote the ground state manifold of $H_0$ as $\mathcal{S}_0$ with the corresponding projector $P$, and treat the rest of the states as belonging to the high-energy subspace $\mathcal{S}_1$ with the corresponding projector $P_1 = 1-P$. In the following we assume that the gap separating the high- and low-energy states of $H_0$ are much larger than the energy scale of the perturbation $\lambda H_1$. The full Hamiltonian can be rewritten as 
\begin{equation}
H =     
\left(\begin{array}{@{}c|c@{}}
E_0 P
  &  \lambda P  H_1 P_1 \\
\hline
   \lambda P_1 H_1 P &
  P_1 (H_0 +\lambda H_1)P_1
\end{array}\right).
\end{equation}
The goal of SW transformation is to find a basis transformation $U: |\psi\rangle \to |\tilde{\psi}\rangle =  U|\psi\rangle$, such that in the new basis the Hamiltonian $\tilde{H} = U H U^\dagger$ is block-diagonal:
\begin{equation}
\tilde{H} = U H U^\dagger =
\left(\begin{array}{@{}c c@{}}
P H_{\text{eff}} P
  &  0 \\
   0 &
   P_1(\tilde{H}_0 + \lambda \tilde{H}_1)P_1
\end{array}\right),
\end{equation}
where $H_{\text{eff}}$ is the effective Hamiltonian defined on the low-energy subspace $\mathcal{S}_0$. In the perturbative regime, the spectrum of $P H_{\text{eff}} P$ forms the low-energy spectrum of $H$ and the eigenstates of $H_{\text{eff}}$ are adiabatically connected to those of $H_0$. 

In practice, we require that the off-diagonal terms vanish up to certain order $\lambda^{N_0}$.  Specifically, we parametrize the unitary matrix as $U = e^{S}$, where $S=\sum_{n=1}^{\infty} \lambda^n S_n$ an anti-hermitian matrix. 
Expanding the transformed Hamiltonian as a series in $\lambda$, we obtain
\begin{equation}\label{eq:H_tilde}
\begin{aligned}
    \tilde{H} &= e^{S} (H_0 + \lambda H_1) e^{-S} = e^{[S,\ \cdot \ ]} (H_0+\lambda H_1)\\
    &=H_0+\lambda H_1 + [S, H_0 + \lambda H_1] + \frac{1}{2!}[S, [S, H_0 + \lambda H_1]]+\frac{1}{3!}[S, [S, [S, H_0 + \lambda H_1]]]+\cdots\\
    &=H_0+\lambda([S_1,H_0]+H_1)
    +\lambda^2\left([S_2,H_0]+[S_1,H_1]+\frac1{2!}[S_1,[S_1,H_0]]\right) +\cdots.
\end{aligned}
\end{equation}
We require the transformed Hamiltonian to be block-diagonal up to $\lambda^{N_0}$, which is equivalent to the equation $[P, \tilde{H}] = 0$ up to $\lambda^{N_0}$. This equation can be solved recursively. To achieve this, we define the quantity 
\begin{equation}
X_n = \frac1{n!}(\partial_\lambda^nH^{\mathrm{eff}})_{\lambda\to0}-[S_n,H_0],
\end{equation}
which represents the coefficient of $\lambda^n$ in the expansion of $\tilde{H}$, minus the commutator $[S_n, H_0]$.
For example, at first and second orders, we have $X_1 = H_1$ and $X_2 = [S_1, H_1] + \frac{1}{2!}[S_1, [S_1, H_0]]$.
With $X_n$, we rewrite the commutative condition $[P, \tilde{H}] = 0$ as
\begin{equation}
    [P, \tilde{H}] = \sum_{n=1}^{\infty} \lambda^n [P, [S_n, H_0] + X_n] = 0,
\end{equation}
which admits the following recursive solution
\begin{equation}\label{eq:recursive_relation}
    S_n= P X_n D-DX_nP,\quad D=-\frac{1-P}{H_0-E_0}.
\end{equation}
Note that $X_n$ is a function of $S_m$ with $m\in \{1,\cdots, n-1\}$. 
The final effective Hamiltonian $H_{\text{eff}}$ of the low-energy subspace is obtained from the projection $\hat{H}_{\text{eff}} = P \tilde{H} P$. 
Here we list the expression of $H_{\text{eff}} = \sum_{n=0}^{N_0} \lambda^n H_{\text{eff}}^{(n)}$ for $N_0=4$:
\begin{align}
\hat{H}_{\text{eff}}^{(0)} &= P H_0 P,\\
\hat{H}_{\text{eff}}^{(1)} &= P H_1 P,\\
\hat{H}_{\text{eff}}^{(2)} &= P H_1 D H_1 P,\\
\hat{H}_{\text{eff}}^{(3)} &= P H_1 D H_1 D H_1 P -\frac{1}{2} P H_1 P H_1 D^2 H_1 P -\frac{1}{2} P H_1 D^2 H_1 P H_1 P, \\
\hat{H}_{\text{eff}}^{(4)} &= P H_1 D H_1 D H_1 D H_1 P-\frac{1}{2} P H_1 P H_1 D^2 H_1 D H_1 P -\frac{1}{2} P H_1 D^2 H_1 P H_1 D H_1 P \\
& -\frac{1}{2} P H_1 P H_1 D H_1 D^2 H_1 P 
-\frac{1}{2}P H_1 D H_1 P H_1 D^2 H_1 P +\frac{1}{2} P H_1 P H_1 P H_1 D^3 H_1 P\\
&-\frac{1}{2} P H_1 D^2 H_1 D H_1 P H_1 P -\frac{1}{2} P H_1 D H_1 D^2 H_1 P H_1 P +\frac{1}{2} P H_1 D^3 H_1 P H_1 P H_1 P.
\end{align}
Note that besides ``connected" terms $P (H_1 D)^{n-1} H_1 P$, there are additional ``disconnected" terms that are composed of connected pieces.

We then apply SW transformation to the model defined in the main text.
In the strong-interaction limit, we divide the Hamiltonian into
\begin{equation}
\begin{aligned}
    H_0 = V_1 \sum_{\langle ij \rangle} \hat{n}_i \hat{n}_j
        + V_2 \sum_{\langle ij \rangle_2} \hat{n}_i \hat{n}_j
        + V_3 \sum_{\langle ij \rangle_3} \hat{n}_i \hat{n}_j,\quad
    \lambda H_1 = -\sum_{ij} t_{ij} \hat{c}_i^\dagger \hat{c}_j,
\end{aligned}
\end{equation}
where $V_1 = 2V_2 = 2V_3$.
The ground state manifold $\mathcal{S}_0$ of the cluster-charging interaction $H_0$ at $\nu=1/3$ is extensively degenerate \cite{zhang2022fractional}, where the ground state configurations satisfy the constraint that each plaquette of the honeycomb lattice is occupied by one fermion. The effective Hamiltonian $\hat{H}_{\text{eff}}$ is interpreted as the virtual hopping processes that connect ground state configurations in $\mathcal{S}_0$. The leading-order contribution to the off-diagonal terms of $\hat{H}_{\text{eff}}$ comes from the eighth order lemniscate operator $\mathcal{L}$ \cite{Mao2023}, as shown in Fig.~\ref{fig:processes}(a). 
Below eighth order, there are processes that renormalize the energies of the ground states, i.e., those that contribute to the diagonal part of $\hat{H}_{\text{eff}}$. In general, they can be written as 
\begin{equation}
   H_{\text{eff}}^{\text{diag}} = \sum_{ij} V_{ij} \hat{n}_i \hat{n}_j + \sum_{ijk} V_{ijk} \hat{n}_i \hat{n}_j \hat{n}_k + \cdots.
\end{equation}

Below, we derive the analytical expressions for $H_{\text{eff}}^{\text{diag}}$ up to fourth order. As we will show later,
the fourth order effective Hamiltonian suffices to lift the degeneracy of the ground state manifold and thus dominates over the eighth-order Lemniscate operator.
For convenience, we can view each fermion as a trimer \cite{Mao2023,zhang2024}, with its vertices located on the dual triangular lattice. The cluster-charging energy of a given configuration is proportional to the number of overlaps between vertices. The manifold $\mathcal{S}_0$ then correspond to all possible non-overlapping trimer coverings, such that there is exactly one trimer vertex per plaquette.
In this picture, it is straightforward to see that $\sum_{n=0}^3 \lambda^n H_{\text{eff}}^{(n)\text{diag}} \propto \mathds{1}$. This is because the virtual hopping processes with length $n\leq3$ only involve a single trimer hopping in a uniform density of vertices, and their contributions to $H_{\text{eff}}^{\text{diag}}$ are independent of the trimer coverings. Therefore, they result in the same energy shift for every ground state in $\mathcal{S}_0$. However, note that the trimer picture does not account for the Pauli exclusion principle, i.e., two fermions can not occupy the same site.
Although the trimer picture suggests that the single-particle process gives the same energy shift up to arbitrary orders, some of the high-order processes are blocked due to the presence of other fermions. For $n\leq 3$, the blocking is irrelevant, as those processes can only move a single fermion within a single plaquette where no other fermion is present because of the cluster-charging constraint. By the same reasoning, 
the term $-\frac{1}{2} P H_1 D^2 H_1 P H_1 D H_1 P -\frac{1}{2} P H_1 D H_1 P H_1 D^2 H_1 P$ in $H_{\text{eff}}^{(4)}$ is also proportional to identity. Terms containing $PH_1P$ vanish because a single move of one trimer would necessarily induce non-zero overlaps, resulting in configurations does not belong to $\mathcal{S}_0$. Thus, the only non-trivial contribution up to fourth order is $P H_1 D H_1 D H_1 D H_1 P$, which involves at most two particles and leads to two-body interactions of the form $\sum_{ij} V_{ij} \hat{n}_i \hat{n}_j$. 

\begin{figure}[htb]
    \centering
    \includegraphics[width=.7\columnwidth]{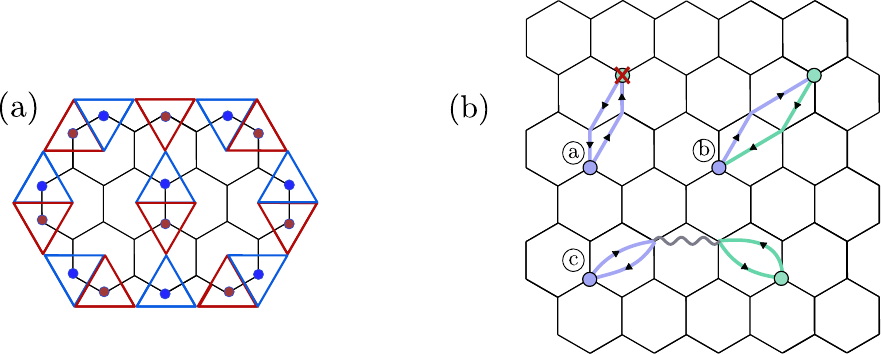}
    \caption{(a) Lemniscate operator. The process maps the fermions from blue sites to red sites, which takes at least eight steps of hopping. (b) Three types of fourth order processes. $\textcircled{a}$ represents a single-particle process that is blocked by the 
    presence of another fermion. $\textcircled{b}$ represents a two-particle process that exchange the two fermions.
    $\textcircled{c}$ represents a two-particle process without exchange, where two fermions directly interact (wavy line) 
    in the intermediate state.}
    \label{fig:processes}
\end{figure}

We then classify the fourth order processes into three types: 
(a) Single-particle processes: As explained earlier, the presence of another fermion will block certain single-particle processes, thereby inducing effective interaction between fermions. 
(b) Exchange processes: These hopping processes exchange two fermions. (c) Virtual-interacting processes: These are two-particle processes without exchanging two particles, but the two moved fermions directly interact through the cluster-charging interaction in the intermediate state. Note that if the two fermions do not interact in the intermediate state, the resulting contributions are independent of the overall fermion configurations, as can be seen from the trimer picture, and thus yields the same energy shift for all ground states.
Figure~\ref{fig:processes}(b) provides illustrations for each type of the process. The induced two-body interactions spans from 4NN to 13NN, which have the form
{{\allowdisplaybreaks
\begin{align}
    V_4: &\quad V_{\text{single}} = \frac{4 t_{1}^{2} t_{2}^{2} \cos^{2}{\left(\varphi \right)}}{3 V_{2}^{3}} + \frac{t_{1}^{2} t_{2}^{2}}{6 V_{2}^{3}} + \frac{2 t_{1} t_{2}^{2} t_{3} \cos^{2}{\left(\varphi \right)}}{V_{2}^{3}} + \frac{2 t_{2}^{2} t_{3}^{2} \cos^{2}{\left(\varphi \right)}}{3 V_{2}^{3}},
    \\
    &\quad V_{\text{exchange}} = \frac{2 t_{1}^{2} t_{2}^{2} \cos{\left(2 \varphi \right)}}{V_{2}^{3}} + \frac{3 t_{1} t_{2}^{2} t_{3} \cos{\left(2 \varphi \right)}}{V_{2}^{3}} + \frac{2 t_{1} t_{2}^{2} t_{3}}{V_{2}^{3}} + \frac{t_{2}^{2} t_{3}^{2} \cos{\left(2 \varphi \right)}}{V_{2}^{3}},\\
    &\quad V_{\text{virtual}} = - \frac{4 t_{1}^{4}}{V_{2}^{3}} - \frac{15 t_{1}^{2} t_{2}^{2}}{4 V_{2}^{3}} - \frac{9 t_{1}^{2} t_{3}^{2}}{4 V_{2}^{3}} - \frac{12 t_{2}^{4}}{5 V_{2}^{3}} - \frac{7 t_{2}^{2} t_{3}^{2}}{5 V_{2}^{3}} - \frac{19 t_{3}^{4}}{60 V_{2}^{3}}, \\
    V_5:&\quad V_{\text{single}} = \frac{5 t_{1}^{2} t_{3}^{2}}{12 V_{2}^{3}} + \frac{t_{1} t_{2}^{2} t_{3}}{V_{2}^{3}} + \frac{2 t_{2}^{4}}{3 V_{2}^{3}},\\
    &\quad V_{\text{exchange}} = \frac{2 t_{1} t_{2}^{2} t_{3}}{V_{2}^{3}} + \frac{2 t_{2}^{4}}{3 V_{2}^{3}}, \\
    &\quad V_{\text{virtual}} = - \frac{2 t_{1}^{4}}{V_{2}^{3}} - \frac{21 t_{1}^{2} t_{2}^{2}}{8 V_{2}^{3}} - \frac{3 t_{1}^{2} t_{3}^{2}}{8 V_{2}^{3}} - \frac{2 t_{2}^{4}}{3 V_{2}^{3}} - \frac{13 t_{2}^{2} t_{3}^{2}}{10 V_{2}^{3}} - \frac{5 t_{3}^{4}}{12 V_{2}^{3}},\\
    V_6: &\quad V_{\text{single}} = \frac{t_{2}^{4}}{6 V_{2}^{3}} + \frac{t_{2}^{2} t_{3}^{2} \cos{\left(2 \varphi \right)}}{3 V_{2}^{3}} + \frac{t_{3}^{4}}{6 V_{2}^{3}}, \\
    &\quad V_{\text{exchange}} = \frac{t_{2}^{2} t_{3}^{2} \cos{\left(2 \varphi \right)}}{2 V_{2}^{3}},\\
    &\quad V_{\text{virtual}} = \frac{2 t_{1}^{4}}{3 V_{2}^{3}} - \frac{3 t_{1}^{2} t_{3}^{2}}{4 V_{2}^{3}} - \frac{3 t_{2}^{4}}{20 V_{2}^{3}} - \frac{9 t_{2}^{2} t_{3}^{2}}{10 V_{2}^{3}} - \frac{t_{3}^{4}}{12 V_{2}^{3}},\\
    V_7: &\quad V_{\text{single}} = \frac{2 t_{2}^{2} t_{3}^{2} \cos^{2}{\left(\varphi \right)}}{3 V_{2}^{3}},\\
    &\quad V_{\text{exchange}} = \frac{t_{2}^{2} t_{3}^{2} \cos{\left(2 \varphi \right)}}{2 V_{2}^{3}}\\
    &\quad V_{\text{virtual}} = - \frac{3 t_{1}^{2} t_{2}^{2}}{8 V_{2}^{3}} - \frac{3 t_{1}^{2} t_{3}^{2}}{8 V_{2}^{3}} - \frac{7 t_{2}^{4}}{15 V_{2}^{3}} - \frac{t_{2}^{2} t_{3}^{2}}{3 V_{2}^{3}} - \frac{t_{3}^{4}}{3 V_{2}^{3}},\\
    V_8: &\quad V_{\text{single}} = V_{\text{exchange}} = 0, \\
    &\quad V_{\text{virtual}} = \frac{2 t_{1}^{4}}{3 V_{2}^{3}} + \frac{3 t_{1}^{2} t_{2}^{2}}{4 V_{2}^{3}} + \frac{4 t_{2}^{4}}{15 V_{2}^{3}} + \frac{t_{2}^{2} t_{3}^{2}}{5 V_{2}^{3}} + \frac{t_{3}^{4}}{10 V_{2}^{3}}, \\
    V_9: &\quad V_{\text{single}} = V_{\text{exchange}} = 0, \\
    &\quad V_{\text{virtual}} = \frac{3 t_{1}^{2} t_{2}^{2}}{8 V_{2}^{3}} + \frac{3 t_{1}^{2} t_{3}^{2}}{8 V_{2}^{3}} + \frac{11 t_{2}^{4}}{60 V_{2}^{3}} + \frac{t_{2}^{2} t_{3}^{2}}{5 V_{2}^{3}}, \\
    V_{10}: &\quad V_{\text{single}} = V_{\text{exchange}} = 0,\\
    &\quad V_{\text{virtual}} = \frac{3 t_{1}^{2} t_{2}^{2}}{16 V_{2}^{3}} + \frac{3 t_{1}^{2} t_{3}^{2}}{16 V_{2}^{3}} + \frac{3 t_{2}^{4}}{20 V_{2}^{3}} + \frac{11 t_{2}^{2} t_{3}^{2}}{60 V_{2}^{3}} + \frac{t_{3}^{4}}{20 V_{2}^{3}},\\
    V_{11}:&\quad V_{\text{single}} = V_{\text{exchange}} = 0, \\
    &\quad V_{\text{virtual}} = \frac{t_{2}^{4}}{10 V_{2}^{3}} + \frac{t_{2}^{2} t_{3}^{2}}{5 V_{2}^{3}} + \frac{t_{3}^{4}}{12 V_{2}^{3}}, \\
    V_{12}: &\quad V_{\text{single}} = V_{\text{exchange}} = 0,\\
    &\quad V_{\text{virtual}} = \frac{t_{2}^{4}}{20 V_{2}^{3}} + \frac{t_{2}^{2} t_{3}^{2}}{10 V_{2}^{3}} + \frac{t_{3}^{4}}{20 V_{2}^{3}}. \\
    V_{13}: &\quad V_{\text{single}} = V_{\text{exchange}} = 0\\
    &\quad V_{\text{virtual}} = \frac{t_{2}^{4}}{20 V_{2}^{3}} + \frac{t_{2}^{2} t_{3}^{2}}{10 V_{2}^{3}} + \frac{t_{3}^{4}}{20 V_{2}^{3}},
\end{align}
}}
The absolute value of $V_{ij}$ decays as $|\mathbf{r}_i - \mathbf{r}_j|$ increases.
Note that $V_{ij}$ have non-trivial dependence on $\varphi$, and so does the ground state. 
In order to determine the ground state of such extended interaction, we assume that the ground state is translationally invariant with respect to some enlarged unit cell. Then we do a exhaustive search on inequivalent enlarged unit cells with number of sites below $48$. For each type of unit cell, we first determine the classical ground states $\mathcal{S}_0$ of $H_0$ and calculate the energy density under $\hat{H}_{\text{eff}}$ for each configuration in $\mathcal{S}_0$, where we take into account both the interactions within a unit cell and the interactions between unit cells. In the end, we pick the unit cell that gives rise to the lowest-energy ground state. As an example, we plot the unit cells of several possible charge order patterns at $\nu=1/3$ in Fig.~\ref{fig:unit_cell_CO}(a): $\sqrt{3}\times\sqrt{3}$, brick wall \cite{Mao2023}, and stripe order. The corresponding energies per fermion are
\begin{align}
    e_{\sqrt{3}\times\sqrt{3}} &= 3 V_5 + 3 V_{12}, \\
    e_{\text{BW}} &= \frac{3}{2}V_4 + V_6 + \frac{1}{2}V_7 + V_{10} + \frac{1}{2} V_{11} + V_{12}, \\
    e_{\text{stripe}} &= V_4 + V_5 + V_7 + V_9 + V_{12},
\end{align}
where $e_{\sqrt{3}\times\sqrt{3}}$ is independent of $\varphi$ and $e_{\text{BW}} > e_{\text{stripe}}$ for all $\varphi$ for the parameters $t_2/t_1 = 0.7, t_3/t_1 = -0.9$ used in the main text.
The exhaustive search reveals two possible ground states, depending on the phase $\varphi$: one is the $\sqrt{3}\times \sqrt{3}$ charge order, and the other is the stripe order, as shown in Fig.~\ref{fig:unit_cell_CO}(b). The transition is driven by the competition among 4NN, 5NN, and 6NN interactions, whose strengths are much larger than the further-range interactions.
In the main text, we fix $\varphi=0.35\pi$, which lies within the $\sqrt{3} \times \sqrt{3}$ gWC phase.

\begin{figure}[htb]
    \centering
    \includegraphics[width=0.6\columnwidth]{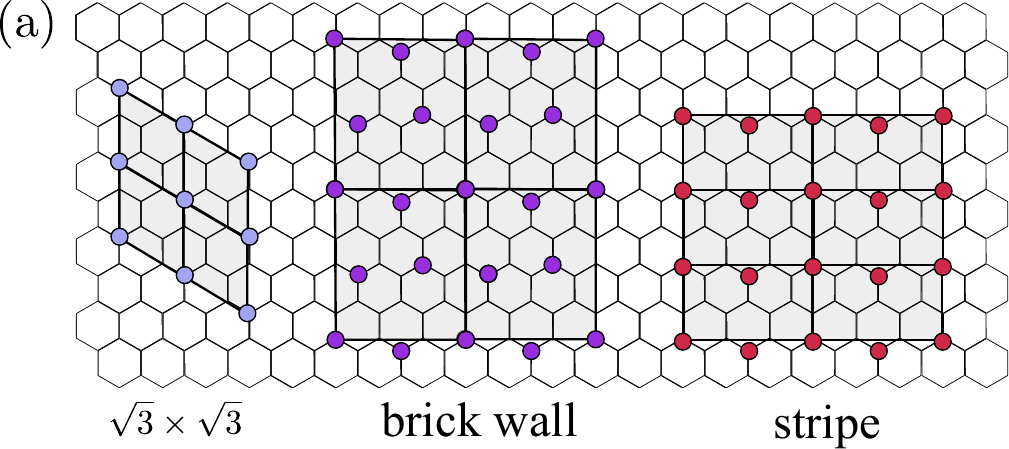}
    \hfill
    \includegraphics[width=0.35\columnwidth]{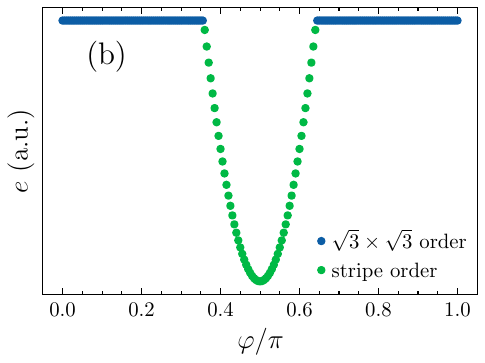}
    \caption{(a) Enlarged unit cells of some typical charge order patterns at $\nu=1/3$ filling. (b) The ground state of
    $\hat{H}_{\text{eff}}^{(4)}$ as the phase $\varphi$ varies. The plot shows the energy density $e$ of the ground state as a function of $\varphi$. The parameters we use are the same as those in the main text: $t_2/t_1 = 0.7$ and $t_3/t_1 = -0.9$.}
    \label{fig:unit_cell_CO}
\end{figure}

\setcounter{figure}{0}
\setcounter{equation}{0}
\renewcommand\thefigure{D\arabic{figure}}
\renewcommand\theequation{D\arabic{equation}}
\section*{Appendix D: Phase diagrams of the constrained model}
\label{app:constrained_model}
In the main text, we introduce a simplified model defined on a constrained Hilbert space, 
which is motivated to capture the essential physics of the original model. 
The Hamiltonian for this simplified model is given by
\begin{equation}
    \hat{H}_{\text{c}}(E_{\text{cutoff}}) = \hat{P}(E_{\text{cutoff}}) \bigg[-\sum_{i, j} t_{ij} \hat{c}_i^\dagger \hat{c}_j\bigg] \hat{P}(E_{\text{cutoff}}),
\end{equation}
where $\hat{P}(E_{\text{cutoff}})$ denotes the projector onto the space with $E \leq E_{\text{cutoff}}$ in the classical cluster-charging model.
In this model, the effect of cluster-charging interaction is incorporated through the projector $\hat{P}(E_{\text{cutoff}})$, which determines the relevant Hilbert space based on the cutoff $E_{\text{cutoff}}$. We study the phase diagrams of $\hat{H}_{\text{c}}$ as a function $E_{\text{cutoff}}$ on finite clusters. As shown in Fig.~\ref{fig:Hc_allsizes}, the phase diagrams of $\hat{H}_{\text{c}}$ on finite clusters exhibit the same phase structures as those of the full Hamiltonian $\hat{H}$, with $E_{\text{cutoff}}$ playing the role of $t_1$. 
Notably, the FCI of $\hat{H}_{\text{c}}$ is generally more prominent, as demonstrated by the large energy separation between the FCI ground states and the excited states. This is illustrated in Fig.~\ref{fig:FCI_24_1}, where we present evidences of the FCI phase for the $(N_1, N_2, \text{deg})=(24, 1, 96)$ cluster with $E_{\text{cutoff}} =2V_1$.
The correspondence between the phase diagrams of $\hat{H}_{\text{c}}$ and $\hat{H}$ justifies the validity of $\hat{H}_{\text{c}}$ and demonstrates that the FCI observed here is distinct from the conventional Laudau-level analogy, where the lower Chern band serves as the relevant Hilbert space.

\begin{figure}[t]
    \centering
    \includegraphics[width=\columnwidth]{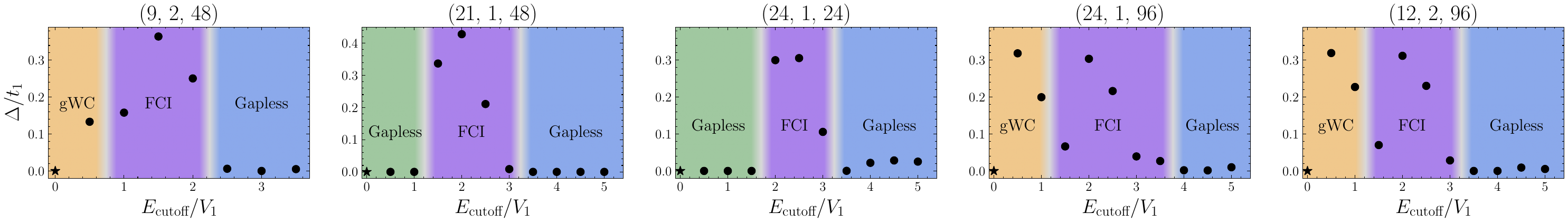}
    \caption{The phase diagrams of the constrained Hamiltonian $H_{\text{c}}$ on finite clusters. The tuple $(N_1, N_2, \text{deg})$ on top of each plot specifies the cluster, whose geometry is shown in the main text. The black data points denote $\Delta/t_1 = (E_4 - E_3)/t_1$.
    The gWC and FCI phases are distinguished by the Chern number calculations. The phase diagrams are qualitatively the same as those of $\hat{H}$, with $E_{\text{cutoff}}$ replacing $t_1$.}
    \label{fig:Hc_allsizes}
\end{figure}

\begin{figure}[h]
    \centering
    \includegraphics[width=\columnwidth]{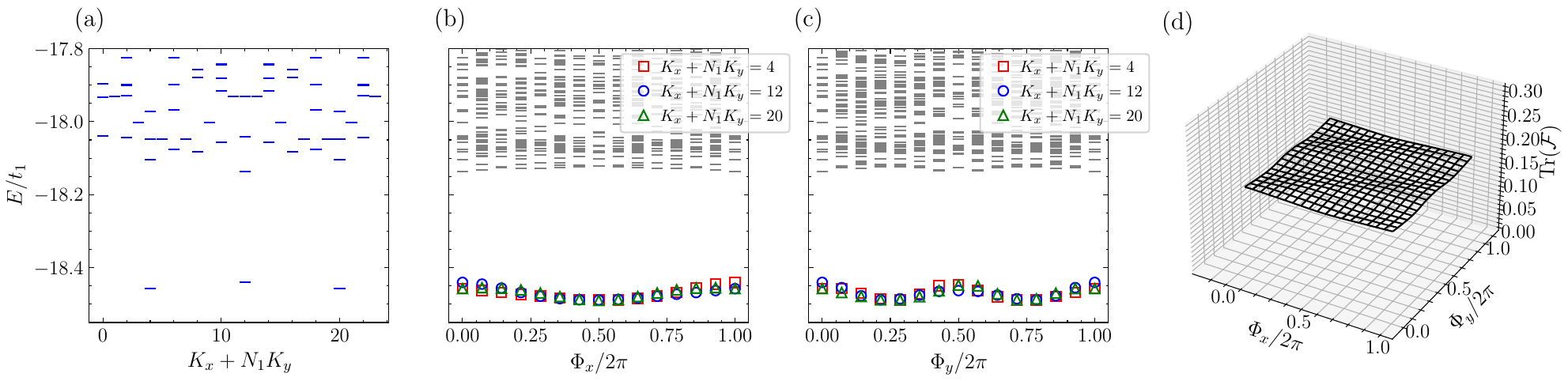}
    \caption{Characterizations of the FCI phase for the $(24, 1, 96)$ cluster at $E_{\text{cutoff}} = 2V_1$. (a) Momentum-resolved energy spectrum. (b, c) Spectral flow upon inserting flux along the $x$- and $y$-directions, respectively. The FCI ground states are marked in non-grey colors. (d) Distribution of the many-body Berry curvature $\text{Tr}(\mathcal{F})$ of the FCI ground states. Integration of $\text{Tr}(\mathcal{F})$ yields $\mathcal{C}=1$.}
    \label{fig:FCI_24_1}
\end{figure}
\end{widetext}

\end{document}